# Navigating the Complex Compositional Landscape of High-Entropy Alloys

Jie Qi[1], Andrew M. Cheung[2], and S. Joseph Poon[1*]

[1] Department of Physics, University of Virginia, Charlottesville, VA 22904-4714

[2] Ph.D. in Engineering Physics, University of Virginia, VA 22904-4259

**Abstract:** High-entropy alloys, which exist in the high-dimensional composition space, provide enormous unique opportunities for realizing unprecedented structural and functional properties. A fundamental challenge, however, lies in how to predict the specific alloy phases and desirable properties accurately. This review article provides an overview of the data-driven methods published to date to tackle this exponentially hard problem of designing high-entropy alloys. Various utilizations of empirical parameters, first-principles and thermodynamic calculations, statistical methods, and machine learning are described. In an alternative method, the effectiveness of using phenomenological features and data-inspired adaptive features in the prediction of the high-entropy solid solution phases and intermetallic alloy composites is demonstrated. The prospect of high-entropy alloys as a new class of functional materials with improved properties is featured in light of entropic effects. The successes, challenges, and limitations of the current high-entropy alloys design are discussed, and some plausible future directions are presented.

* Corresponding author: sjp9x@virginia.edu

**Table of content:**

## Section 1: Introduction

High-Entropy Alloys (HEAs), which reside in a complex composition space, are stabilized to an extent by their high entropy of mixing to form solid-solution alloys with normally unattainable compositions [1–8]. The axes of the high-dimensional composition space, in which the HEAs exist, are the contents of all the possible elements used to make HEAs. The vast compositional flexibility of HEAs provides a new concept for discovering desirable properties. In fact, HEAs have been shown to exhibit some outstanding mechanical properties thanks to a superior balance of strength and fracture toughness, surpassing those of conventional alloys [4–6, 8–18]. Besides mechanical properties, studies that focus on the physical and functional properties of HEAs have also begun to emerge [19]. Several properties, most notably corrosion-resistant [20–22], electrocatalytic [23, 24], thermal-resistant [25, 26], and thermoelectric properties [27, 28], were found to be comparable to or better than those of conventional alloys. To date, only very limited regions of the compositional space have been investigated.

The challenges in exploiting the nearly inexhaustible opportunities for designing HEA materials with superior structural and functional properties are both empowering and daunting. Since different types of properties have different characteristic behaviors, the design of high-performance HEAs must involve various specific constituent elements and compositions, as well as lattice structures and microstructures. The complexity of the exponentially large design space presents a great challenge in the prediction of specific compositions and phases. Significant scientific undertaking and development are needed in order to capitalize on the vast discoverable opportunities offered by the high-dimensional composition space of HEAs.

The formation of high-entropy phases is primarily controlled by thermodynamic and kinetic factors. For understanding the growing number of HEAs, empirical methods that utilized atomistic and thermodynamic parameters were introduced to investigate HEA compositional regions [8, 29, 30]. The empirical approaches were later complemented by first-principles calculations [31] and Calculation of Phase Diagrams (CALPHAD) [32, 33] to shed light on the thermodynamic factor of HEA formation. Monte Carlo simulations showed promising results in predicting the formation of intermetallic phases and the evolution of phase structures with temperature [34]. Despite progress in understanding the formation trend of HEAs, much of the alloy design for HEAs still relies on trial and error experimentation. Recently, there have been increasing efforts in employing data-driven methods to exploit the growing data set of HEAs. Some initial methods included the utilization of statistical models complemented with thermodynamics [35], and as well as high-throughput (HTP) experimentation [36, 37] designed to underpin the HEA phase formation trend. The thin film deposition method employed in the HTP study tended to result in the metastable HEA phases. Meanwhile, the utilization of machine learning (ML) models has demonstrated some initial promise in phase prediction and property design (e.g., high hardness) [38, 39]. Most ML has focused on supervised learning with different models, such as the support vector machine (SVM) [38–43] and artificial neural network (ANN) [40, 43, 44]. The ML models were trained using atomistic and thermodynamic parameters. Despite some success in categorizing the compositional regions of certain solid solution (SS) and intermetallic (IM) phases, the predictions often fell short of differentiating between the specific phases. On the other hand, a ML model that utilized phenomenological features obtained from binary phase diagrams was found to achieve high accuracy in categorizing specific phase formation [45].

The various computation and experimentation methodologies will continue to develop towards advancing the science and design of high-entropy alloys. In view of this promising development, this review serves as a timely report of the status of progress in harnessing the high-dimensional composition space of HEAs as a requisite for designing the desired properties. Section 2 provides a comprehensive review of the various methods for predicting the occurrence of HEA phases in the complex composition space. Section 3 introduces an alternative machine learning model using phenomenological features for predicting HEA phase formation. Section 4 discusses the potential of HEAs as a new class of high-performance structural and functional alloys in light of the core effects of HEAs. Section 5 highlights some critical scientific issues along with plausible future developments. Section 6 is the conclusion.

## Section 2: Methods for Predicting High-Entropy Phase Formation

The compositional and phase complexities of HEAs required the use of phase prediction models to economize the time and effort needed to synthesis new alloys. Traditionally, inefficient trial and error design approaches were used. Here, different efforts that aim to circumnavigate the traditional approaches are presented. In this section, phase predictive approaches that use empirical methods, first-principles calculations, and statistical studies utilizing ML are reviewed.

Common SS phases found in HEAs include FCC (face-centered cubic), BCC (body-centered cubic phase), and HCP (hexagonal close-packed). The Strukturbericht designations for the disordered and the ordered phases of these SS phases are A1 and $L1_2$, A2 and B2, or A3 and $D0_{19}$, respectively. Common IM phases appearing in HEAs are Sigma phases ($D8_b$) with tetragonal crystal structure; Laves phase with cubic (C14), hexagonal (C15), or hexagonal (C36) structure; and μ phases ($D8_5$) with rhombohedral structure.

### 2.1 Empirical methods

The initial methods were based on empirical parameters. They incorporate aspects of the minimization of Gibbs Free Energy, Hume-Rothery theory, electronic configuration, and lattice strain to form phase formation rules. These empirical parameters, defined and discussed in sections 2.1.1-3, were used to either individually or in conjunction determine potential HEA candidates. The correlations between these parameters and the HEA phases formed are analyzed in section 2.1.4.

### 2.1.1 Free energy parameters

The phases favored during the solidification of a HEA possess the lowest Gibbs free energy of mixing ($\Delta G_{mix}$). Parameters associated with $\Delta G_{mix}$ are the mixing entropy ($\Delta S_{mix}$), the mixing enthalpy ($\Delta H_{mix}$), and labeled parameters $\Omega$, $\phi$, $\Phi$, $\eta$, and $k_1^{cr}$, defined in Eqn. 2-8. $\Delta G_{mix}$ is defined as

$$\Delta G_{mix} = \Delta H_{mix} - T\Delta S_{mix} \quad (\text{Eqn. } 1)$$

where T is the phase formation temperature. The $\Delta S_{mix}$ for forming a single or multiple SS phase is approximated as the configurational entropy [2] ($\Delta S_C$) and is calculated according to Boltzmann's hypothesis (Eqn. 2):

$$\Delta S_{mix} \approx \Delta S_C = -R\sum_{i=1}^{N} c_i \ln(c_i) \quad (\text{Eqn. } 2)$$

where R is the gas constant and $c_i$ is the atomic percentage of the i-th element for a N-component system. The definitions of $c_i$ and N are the same throughout this chapter. A SS phase formation is energetically favored over an IM phase formation when the $\Delta S_{mix}$ term is larger.

The $\Delta H_{mix}$ term represents the chemical compatibility among the elements in HEAs [46]. For

HEAs, the $\Delta H_{mix}$ for forming a SS phase is typically calculated from Miedema's model [47] (Eqn. 3):

$$\Delta H_{mix} = \sum_{i=1, i \neq j}^{N} 4\ \Delta H_{I,j}^{mix}\ c_i c_j \quad (\text{Eqn. } 3)$$

where $\Delta H_{I,j}^{mix}$ is the binary mixing enthalpy of an i-j elemental pair. An increase in the negativity of $\Delta H_{mix}$ increases the probability of forming an IM phase. A positive $\Delta H_{mix}$ indicates an immiscibility among certain elements, which could lead to phase separation. As shown in Fig. 3(a), $-16\frac{kJ}{mol} < \Delta H_{mix} < +5\frac{kJ}{mol}$ is the criterion proposed for forming a single SS phase [29].

$\Delta G_{mix}$ is determined by the entropy and enthalpy terms. Whether SS or IM phase formation is favored is dependent on the interplay of these two terms. The parameters $\Omega$ and $\phi$ are used to compare the magnitudes of the entropy and enthalpy terms. Zhang et al. [48] defined the $\Omega$-parameter as

$$\Omega\ = \frac{T_m \Delta S_{mix}}{|\Delta H_{mix}|} \quad (\text{Eqn. } 4)$$

where $T_m = \sum_{i=1}^{N} c_i\ T_{m_i}$ is the HEA melting temperature and $T_{m_i}$ is the melting temperature of the i-th element. A large $T_m \Delta S_{mix}$ or $|\Delta H_{mix}|$ term stabilizes SS or IM phase formation, respectively. When $\Omega\ >\ 1$ [48], as shown in Fig. 3(b), a SS phase formation is favored.

Ye et al. [49] defined the $\phi$-parameter. The total configurational entropy of mixing ($\Delta S_T$) is defined as $\Delta S_T = \Delta S_C +\ \Delta S_E$, where $S_C$ is the configurational entropy of mixing for an ideal gas and $\Delta S_E$ is the excessive entropy of mixing [50]. $\Delta S_T$ deviates from the approximation of $\Delta S_T \approx \Delta S_C$ due to the influence of factors such as differences in atomic size and the packing fraction. $\Delta S_E$, usually negative, is introduced to represent this deviation and $\Delta S_T$ is adjusted by its absolute magnitude, $\Delta S_T = \Delta S_C - |\Delta S_E|$. The parameter $\phi$ defined as

$$\phi = \frac{\Delta S_C - \left|\frac{\Delta H_{mix}}{T_m}\right|}{|\Delta S_E|} > 1 \quad (\text{Eqn. } 5)$$

is the result of combining $\Delta S_T$ and $\Omega$. Based on existing values of known HEAs, shown in Fig. 3(c), $\phi\ > 7$ is the proposed range for SS phase formation [29].

Instead of only comparing the enthalpy and entropy terms for predicting the formation of SS or IM phases, the parameters $\Phi$, $\eta$, and $k_1^{cr}$ were defined by examining difference in $\Delta G_{mix}$. King et al. [51] defined the $\Phi$-parameter to compare the $\Delta G_{mix}$ for forming a fully disordered SS phase ($\Delta G_{SS}$) with the $\Delta G_{mix}$ for IM formation or phase segregation ($\Delta G_{max}$). The $\Phi$-parameter is defined as

$$\Phi = \frac{\Delta G_{SS}}{-|G_{max}|} \quad (\text{Eqn. } 6)$$

where $|\Delta G_{max}|$ represents the absolute magnitude of the larger of the following two values: the lowest possible negative $\Delta G_{mix}$ when the strongest binary compound forms, or the highest possible positive $\Delta G_{mix}$ when a phase is segregated due to the positive mixing enthalpy between two specific constituent elements. When $\Phi\ >\ 1$, then SS phase formation is favored.

Troparevsky et al. [52] defined a parameter, later labeled by others as $\eta$, that is a first order approximation used to compare the $\Delta G_{mix}$ for forming SS and IM phases. The $\Delta H_{mix}$ for SS phase formation and the $\Delta S_{mix}$ for forming IM phases are usually small. Thus, the entropy contribution $-T_{ann}\Delta S_{mix}$, where $T_{ann}$ is the annealing temperature of a HEA, is used to approximate the $\Delta G_{mix}$ for forming SS phases. The enthalpy of formation ($\Delta H_f$), the most negative binary mixing enthalpy for IM phase formation among the constituent element pairs derived from density functional theory (DFT) calculations, is used to approximate the $\Delta G_{mix}$ for forming IM phases. $\eta$ is then defined as

$$\eta = \frac{-T_{ann}\Delta S_{mix}}{|\Delta H_f|} \quad \text{(Eqn. 7)}$$

where an increasingly larger value of the parameter indicates a favorability for forming a SS phase. $\eta > 0.19$ [29] is the proposed lower boundary for the region of SS phase formation as seen in Fig. 3(c).

Similar to the formulation of $\eta$, Senkov and Miracle [53] developed parameters to compare the $\Delta G_{mix}$ for forming SS and IM phases. Their approach was less approximate. Their criterion for forming SS phase is $\Delta H_{mix} - T\Delta S_{mix} < \Delta H_{IM} - T\Delta S_{IM}$, where $\Delta H_{mix}$ and $\Delta H_{IM}$ are the mixing enthalpies, and $\Delta S_{mix}$ and $\Delta S_{IM}$ are the mixing entropies for forming SS and IM phases, respectively. $\Delta S_{IM}$ for the IM phase is approximated to be $0.6\Delta S_{mix}$. This relation of a simple thermodynamic criterion can be expressed as

$$k_1^{cr} = 1 - \frac{0.4T\,\Delta S_{mix}}{\Delta H_{mix}} > \frac{\Delta H_{IM}}{\Delta H_{mix}} \quad \text{(Eqn. 8)}$$

The parameters $k_1^{cr}$ and $\frac{\Delta H_{IM}}{\Delta H_{mix}}$ are plotted in Fig. 3(d). When $k_1^{cr} > \frac{\Delta H_{IM}}{\Delta H_{mix}}$, a SS phase formation is favored.

Pei et al. [54] defined the parameter $\gamma$ to compare the $\Delta G_{mix}$ for forming single phases and multi-phases. For $\gamma$, $\Delta G_{mix} = \Delta H_{mix} - \alpha T_m \Delta S_c$, where $\alpha$ is a scaling parameter. The $\Delta H_{mix}$ definition varies from other methods by using a combination of the formation enthalpy calculated based on the Lennard-Jones potential, and the strain-induced energy calculated from the Kanzaki force [55]. For any given composition, $\Delta H_{mix}$ was calculated for FCC, BCC, HCP, and simple cubic structures. The minimum $\Delta H_{mix}$ was adopted to calculate $\Delta G_{mix}$. For the entropy term, the real system entropy was typically smaller than $\Delta S_c$, and the real temperature when the SS phase was stable could be below $T_m$. Consequently, $\alpha$ is a scale-down factor for the entropy contribution from the ideal to the real conditions. Optimum phase separation occurred at a value of $\alpha = 0.25$. For a N-component HEA, the $\Delta G_{mix}$ was calculated for all the constituent binaries ($\Delta G_2$) and the HEA ($\Delta G_N$). The uniform SS phase formation ability depended on $\Delta G_N$ and the smallest $\Delta G_2$ value, $\min(\Delta G_2)$. Thus, $\gamma$ was defined as

$$\gamma = \begin{cases} \dfrac{\Delta G_N}{\min(\Delta G_2)}, & \text{if } \min(\Delta G_2) < 0; \\ -\dfrac{\Delta G_N}{\min(\Delta G_2)}, & \text{if } \Delta G_N < 0 \text{ and } \min(\Delta G_2) > 0; \end{cases} \quad \text{(Eqn. 9)}$$

The criterion for forming the SS phase was $\gamma \geq 1$. $\gamma$ was used to test 296 existing HEAs in BCC, FCC, HCP, and multi-phases. While 73 % were classified correctly, when jointly using $\gamma$ and the radius mismatch ($\delta < 6$ %, defined in Section 2.1.2), 81 % consistency was obtained, as demonstrated in Fig. 1a. The validity of $\gamma$ was further confirmed by using CALPHAD. $\gamma$ was calculated for each of the 1,146 equimolar HEAs. The compositions were selected, as seen in Fig. 1b, from three 9-element blocks in the periodic table. From these blocks, 266 single SS phases HEAs with 74 BCC, 145 HCP, and 47 FCC phases were obtained. Of the 266 predicted HEAs, only 77 could be validated with CALPHAD due to the limitations of the thermal databases. However, the 77 had a high validation consistency of 94 %.

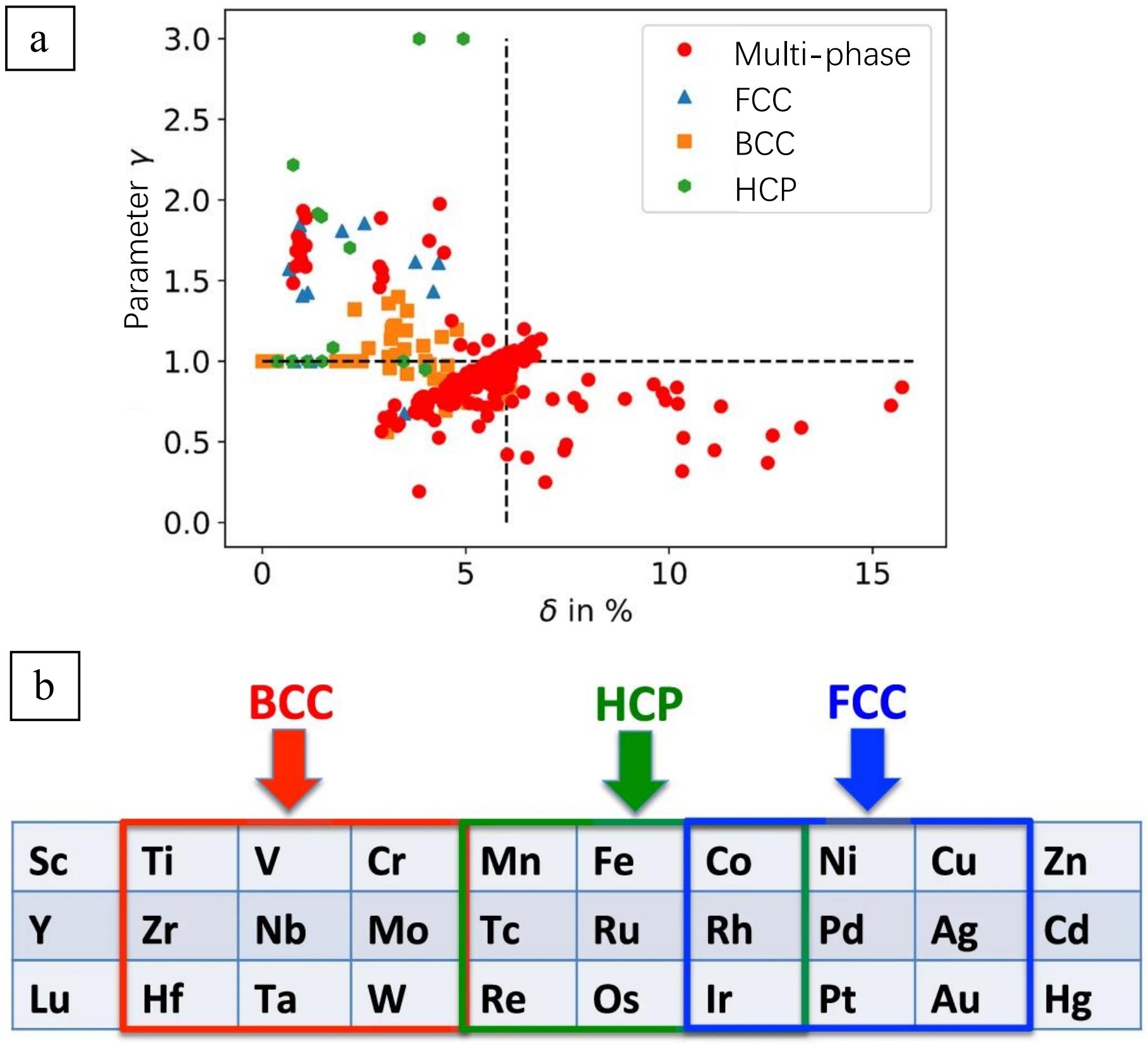

Figure 1. (a) Plot of the distribution of FCC, BCC, HCP, and multi-phase HEAs for parameters $\gamma$ and $\delta$. The criteria for forming a single SS phase are bounded by $\gamma \geq 1$ and $\delta < 6$ %. (b) Equimolar HEAs with BCC, HCP, and FCC predicted phases were obtained from three 9-element blocks in the periodic table. Figures from Pei et al. [54]

### 2.1.2 Parameters from Hume-Rothery theory

According to the Hume-Rothery theory [56], the formation of a SS phase is influenced by radius mismatch, electronegativity mismatch, and electron concentration among the constituent elements. Based on this theory, to study HEAs, several parameters that can influence the phase formation were defined.

The intrinsic residual strain, caused by the radius difference, makes the multi-phase formation possible. Parameters $\delta$, $\sqrt{<\varepsilon^2>}$, $\frac{E_2}{E_0}$, $s_m$, and $K_m$, defined below, relate HEA phase formation to intrinsic strain.

Small radii differences between constituent elements, equivalent to small lattice distortions, favor the formation of the SS phase. The radius mismatch of an alloy ($\delta$) [46] is calculated by

$$\delta = \sqrt{\sum_{i=1}^{N} c_i \left[1 - \frac{r_i}{\sum_{j=1}^{N} c_j\ r_j}\right]^2} \text{ (Eqn. 10)}$$

where $r_i$ is the atomic radius of the i-th element. $\delta < 6$ % [29] is the region for SS phase formation, as seen in Fig. 3(a).

Ye et al. [57] developed a geometric model to calculate the root-mean-square residual strain $\sqrt{<\varepsilon^2>}$ from other parameters such as the atomic percentage, atomic size, and packing density. The mean-square is defined as residual strain $<\varepsilon^2> = \sum_{i=1}^{N} c_i {\varepsilon_i}^2$, where $\varepsilon_i$ is the residual strain of the i-th element in a N-component system. After derivation, $\varepsilon_i$ can be expressed as

$$\varepsilon_i = \frac{\sum_{j=1}^{N} \omega_{ij} c_j}{\sum_{k=1}^{N} A_{ik} c_k} - \frac{4\pi\eta_{ideal}}{N_i \sum_{k=1}^{N} A_{ik} c_k} \quad \text{(Eqn. 11)}$$

where $\omega_{ij}$ is the solid angle subtended by j-th element around the i-th element with $\omega_{ij} = 2\pi\left[1 - \frac{\sqrt{r_i(r_i+2r_j)}}{r_i+2r_j}\right]$, $r_j$ is the atomic radius of the j-th element, $A_{ik}$ is a dimensionless constant with $A_{ik} = \frac{2\pi x_{ik}}{(x_{ik}+1)^2\sqrt{x_{ik}(x_{ik}+2)}}$ for i-th and k-th elements, $\eta_{ideal}$ is the ideal atomic packing fraction and is computed by $\eta_{ideal} = \frac{1}{2}\sum_{i=1}^{N}\sum_{j=1}^{N} c_j c_i N_i \left[1 - \frac{\sqrt{x_{ij}(x_{ij}+2)}}{x_{ij}+1}\right]$, $x_{ij}=r_i/r_j$ is the atomic radius ratio, and $N_i$ is the coordinate number of the i-th atom. A significantly large $\sqrt{<\varepsilon^2>}$ leads to lattice distortions that disrupt single-phase lattices and form multi-phase lattices.

Wang et al. [58] defined another parameter $\frac{E_2}{E_0}$ related to the intrinsic elastic strain energy. In an ideal N-component uniform HEA lattice, the average atomic radius is $\bar{r} = \sum_{i=1}^{N} c_i r_i$. In a real lattice, atoms are displaced from $\bar{r}$. The dimensionless strain is calculated to be $\Delta d = \frac{|r_i+r_j-2\bar{r}|}{2\bar{r}}$. The dimensionless parameter $\frac{E_2}{E_0}$ is defined as

$$\frac{E_2}{E_0} \propto (\Delta d)^2 = \sum_{j\geq i}^{N} \frac{c_i c_j |r_i + r_j - 2\bar{r}|^2}{(2\bar{r})^2} \quad \text{(Eqn. 12)}$$

where a low value of $\frac{E_2}{E_0}$, similar to small values of $\sqrt{<\varepsilon^2>}$, favors the SS phase formation. Fig. 3(e) is a plot of $\frac{E_2}{E_0}$ and $\sqrt{<\varepsilon^2>}$ for HEAs with different phases. It demonstrates that when $\frac{E_2}{E_0} < 13.6\times 10^{-4}$ and $\sqrt{<\varepsilon^2>} < 6.1\ \%$, single-phase HEAs tend to form [29].

Interatomic spacing mismatch ($s_m$) and the bulk modulus mismatch ($K_m$) were developed by Toda-Caraballo et al. [30, 59] with

$$s_m = \sum_{i=1}^{N}\sum_{j=1}^{N} c_i c_j \left|1 - \frac{s_{ij}^d}{s_{lat}}\right| \quad \text{(Eqn. 13)}$$

and

$$K_m = \sum_{i=1}^{N}\sum_{j=1}^{N} c_i c_j \left|1 - \frac{K_{ij}^d}{K_{lat}}\right| \quad \text{(Eqn. 14)}$$

where $s_{ij}^d$ and $K_{ij}^d$ are two matrices representing the interatomic spacing and bulk modulus for i-j atom pairs, $K_{lat}$ is the bulk modulus of the lattice, and $s_{lat}$ is the mean interatomic distance across the lattice. Fig. 2 shows the HEA phase separation based on parameters $s_m$ and $K_m$. In the plot, SS phases tend to from when $s_m$ is small. This result, again, implies that small lattice distortion prompts the SS phase formation. As for the influence of $K_m$, the FCC phase tends to form when $K_m < 4$, while BCC phase forms when $K_m > 4$, implying that the different forces acting on atoms in a FCC lattice are closer to being homogeneous than those acting on a BCC lattice.

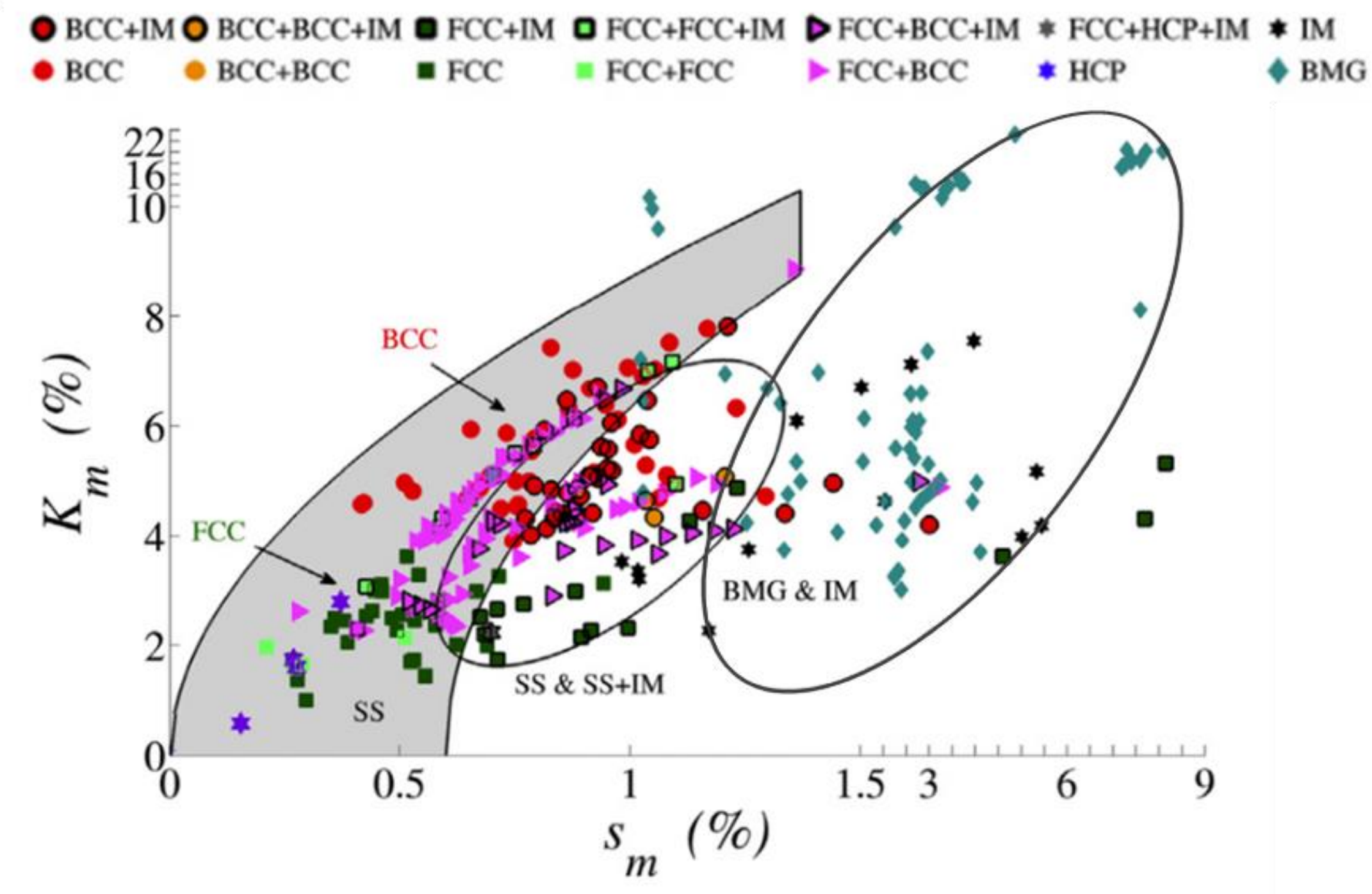

Figure 2. The plot of the distribution of HEA phases for parameter $s_m$ and $K_m$. Figure from Toda-Caraballo and Rivera-Díaz-Del-Castillo [30].

In addition to the effects from the intrinsic strain, electronegativity difference ($\Delta\chi$) and electron configuration are also considered. A small $\Delta\chi$ has been shown to promote SS phase formation[60]. $\Delta\chi$ is defined as

$$\Delta\chi = \sqrt{\sum_{i=1}^{N} c_i \left[\chi_i - \sum_{j=1}^{N} c_j \chi_j\right]^2} \quad \text{(Eqn. 15)}$$

where $\chi_i$ is the i-th HEA element electronegativity. Dong et al. [61] showed that the formation of the Topological Close-Packed (TCP) phases such as Sigma, Laves, and $\mu$ phases can be influenced by $\Delta\chi$ when $\Delta\chi > 0.133$.

Another parameter is the electron concentration, which has two definitions according to Guo et al [62]. The first one is $\frac{e}{a}$ which is the average number of itinerant electrons per atom:

$$\frac{e}{a} = \sum_{i=1}^{N} c_i \left(\frac{e}{a}\right)_i \quad \text{(Eqn. 16)}$$

where $\left(\frac{e}{a}\right)_i$ is the itinerant electrons per atom of the i-th element. The second one is the valence electron concentration (VEC) [62–64] which is the total number of electrons including the *d*-electrons held in the valence band. VEC is defined as

$$\text{VEC} = \sum_{i=1}^{N} c_i \ \text{VEC}_i \quad \text{(Eqn. 17)}$$

where $\text{VEC}_i$ is the VEC of the i-th element. VEC was found to be superior to $\frac{e}{a}$ in predicting HEA

phases. Fig. 3(f) shows that BCC phases form when $\mathrm{VEC} < 6$, FCC phases form when $\mathrm{VEC} > 7.8$, and mixed FCC-BCC phases form when $6 < \mathrm{VEC} < 7.8$ [29]. Tsai et al. [64] applied VEC to a study on $\sigma$ phase formation. They discovered that $6.88 < \mathrm{VEC} < 7.84$ is the $\sigma$-prone formation region for HEAs containing Cr or V.

### 2.1.3 Other parameters

The average value of a d-orbital energy level ($\overline{\mathrm{Md}}$) was proposed by Lu et al. [65] $\overline{\mathrm{Md}}$ is related to the electronegativity and metallic atomic radii. The TCP formation is influenced by $\overline{\mathrm{Md}}$. TCP phases form when $\overline{\mathrm{Md}} > 1.09$ and no TCP formation occurs when $\overline{\mathrm{Md}} < 0.95$.

Poletti et al. [66] proposed the parameter $\mu = T_m/T_{SC}$, where $T_{SC}$ is the spinodal decomposition temperature. A large gap between $T_m$ and $T_{SC}$ can prompt the single SS phase formation at high temperatures. As a result, $\mu > 1.5$ is the proposed region for single SS phase formation.

### 2.1.4 Correlation between the parameters and phase formation

Gao et al.[29] compared the effectiveness of these empirical parameters by coupling and plotting them in Fig. 3. Although the correlation between the parameters and a phase formation exists, precise phase predictions based solely on pairs of these parameters is challenging. Different phases on the plots overlap with ambiguous separation. Additionally, specific phase content in certain categories, such as “multi-phase” and “IM,” were not included. However, these empirical parameters provide fundamental ideas for applying ML to HEA phase formation research. These parameters are related to different aspects of phase formation. Many ML methods utilize these parameters to consider all the phase formation factors comprehensively and this results in improved predictions. Further details are found in section 2.3.

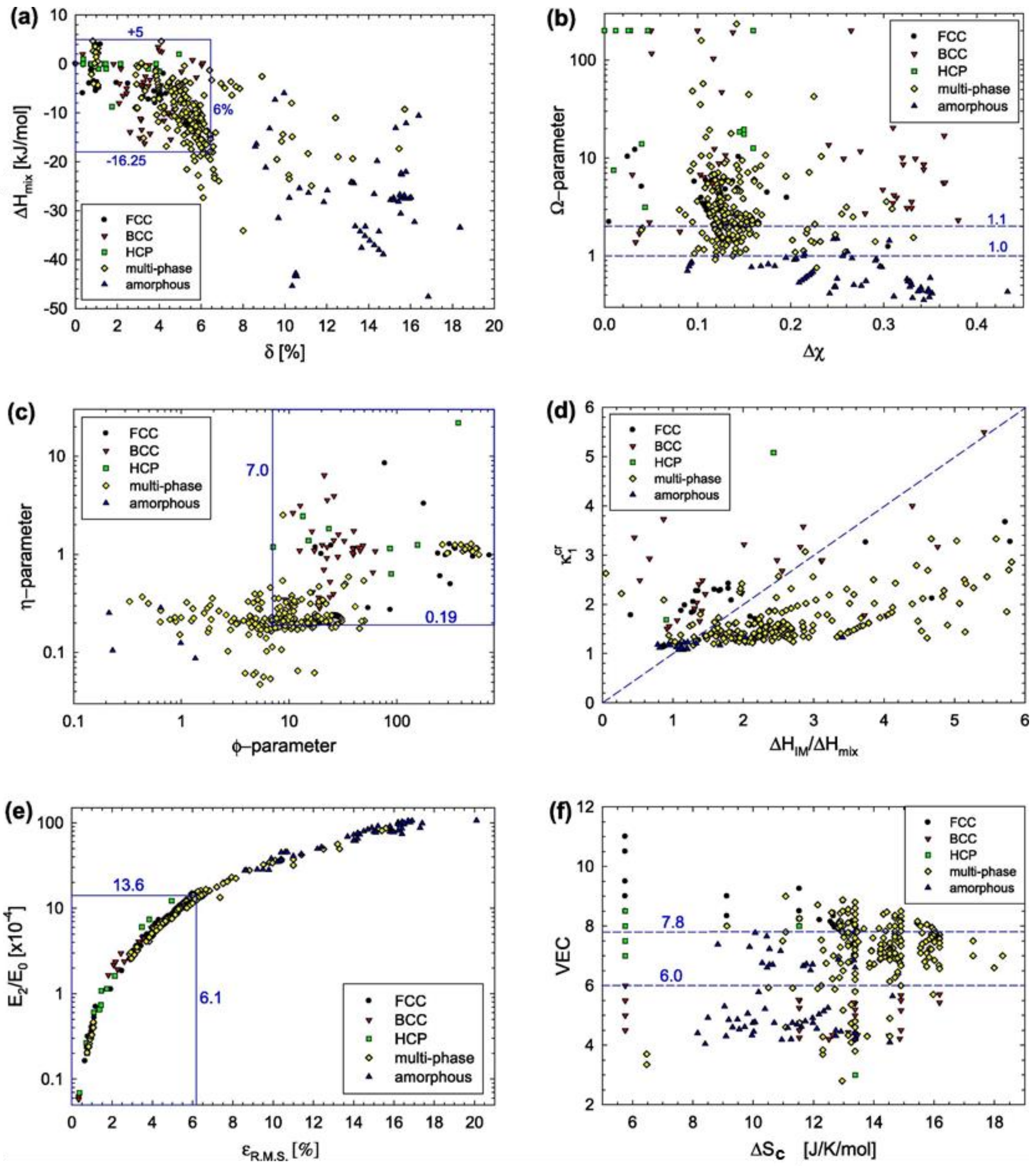

Figure 3. Plots comparing different empirical parameter effects in separating HEA phases. Figure from Gao et al [29].

### 2.2 CALPHAD and first-principles calculations

First-principles calculations provide HEA phase formation results directly from basic physical properties. This section highlights a few studies on HEA phase formation using CALPHAD, ab-initio calculations, DFT, and Monte Carlo (MC) simulations.

#### 2.2.1 CALPHAD

CALPHAD [67–69] is a direct method for determining HEA phase formation. It is a powerful methodology that is employed to predict phase formation and thermodynamic properties such as the composition or temperature boundaries for phase transformations, and precipitation nucleation barrier [70]. Thermodynamic databases, which are the core of CALPHAD, are obtained either from experimental data or DFT results [67]. In order to have accurate HEA phase predictions, the database should ideally include thermodynamic data from lower-order binary or ternary systems [32] that can be extrapolated [71] to simulate higher-order multi-component HEA systems.

As a prediction method that gives detailed phase information, CALPHAD is used widely [20, 32, 33, 63, 67, 72–80] in exploring vast compositional spaces. However, due to the vast compositional space in which HEAs lie, a limitation for the use of CALPHAD is the availability of sufficiently complete binary and ternary thermodynamic databases [32, 81]. Current multi-component alloy databases are designed for traditional alloy systems based primarily on elements such as Al, Fe, Ni and Ti. Complete data for a multitude of ternary systems have yet to be acquired. Without this data, the predictions will not be fully accurate. Additionally, the veracity of the CALPHAD phase predictions drops when miscibility gaps or IM phases are present in the phase diagrams [70].

### 2.2.2 Ab-initio simulations and density functional theory

Ab-initio simulations [31] predict the thermodynamic and mechanical properties [82–92] of HEAs. These properties are determined from simulated electron density which was found using DFT. DFT provides an approximated solution to the Schrodinger equation of a simulated alloy system. An advantage of ab-initio calculations is that they rely on the fundamental quantum mechanical properties of the system, and no experimental nor empirical inputs are needed. However, ab-initio simulations can be computationally intensive methods. Furthermore, in practical use, they are usually used collaboratively with experimental results or other simulation methods like CALPHAD.

DFT calculations can provide binary phase formation information such as the formation energy [52] or bonding strengths [76–78]. Strong binary bonding indicates IM phase formation during solidification. A large positive binary formation energy indicates a potential phase separation. These empirical values can be used as a guide for finding systems with homogeneous binary pairs that tend to form SS phases. However, these methods usually do not distinguish between different SS phases such as the FCC, BCC, or HCP phases.

Ab-initio simulations can calculate $\Delta G_{mix}$ for forming a specific phase. These results have improved accuracy over results using only empirical parameters $\Delta H_{mix}$ and $\Delta S_{mix}$. For a given composition and phase, the contributions to $\Delta G_{mix}$ from the electronic energy, magnetic free energy, atomic vibration free energy, and the configurational entropy can all be computed with certain approximations at various temperatures [93–95]. Ideally, the $\Delta G_{mix}$ selected from the phase formation determination process is the most negative of all possible phase configurations calculated. In practice, determining all the phase configurations is not possible or computationally exorbitant.

Alternatively, specific strategies have been used to expedite the simulation process [31]. The first approach is the combination of ab-initio and existing experimental results. For example, the experiment results from AlCoCrFeNi-type HEAs reveal that the FCC, BCC, and FCC+BCC phases are the phases that can form [96]. Based on these experimental results, Tian et al. [93] used ab-initio calculations to determine the $\Delta G_{mix}$ for forming FCC and BCC phases in AlCoCrFeNi-type HEAs. They then inferred the theoretical compositional space to form FCC, BCC, and mixed FCC+BCC phases. Another approach is enumerating the most probable phases. Wang et al. [97] used this approach to study the phase formation of MoNbTaVW. They computed the $\Delta G_{mix}$ at given temperatures for 178 phases. The calculated $\Delta G_{mix}$ values were analyzed to determine the phase stabilities, phase separation tendencies, and order-disorder transitions. The third approach is studying binary phase diagrams [98, 99]. For example, Rogal et al. [98] from inspection of constituent binary phase diagrams selected the $D0_{19}$ and HCP phases as candidate phases for the HEA $Al_{15}Hf_{25}Sc_{10}Tr_{25}Zr_{25}$. DFT calculations show that the HCP to $D0_{19}$ phase transition occurs at 1230 K, which is in agreement with the experimental result.

Computations of long-range order (LRO) and short-range order (SRO) in HEAs provide valuable information. Ab-initio calculations, coupled with MC or molecular dynamics (MD) simulations, are used for chemical order studies to determine order-disorder transition temperatures [86, 88, 97, 98, 100–111]. Santodonato et al. [34] used MC simulations with inputted DFT results [52] to determine, in the HEA $Al_xCoCrFeNi$ system with variations in Al content, the change in the phase transformation temperature for the precipitation of B2 phase from the BCC phase. Lederer et al. [112] developed a high-throughput ab-initio method to search for potential disordered SS HEAs from 1,240 candidates. The ab-initio results from AFLOW [113], a software framework for high-throughput calculations of crystal structure properties, are then incorporated with a generalized quasi-chemical approximation model [114], generating a temperature-dependent HEA order parameter. The order-disorder transition temperatures were estimated based on the change of this parameter. Furthermore, for a HEA the comparison between the order-disorder transition and melting temperatures was indicative of the predicted disordered SS phase formation tendency. The accuracy of the model was corroborated with MC simulations, experimental data, and CALPHAD, showing high agreement in both SS system predictions and transition temperature predictions.

### 2.3 Statistical and machine learning studies

Since the discovery of HEAs in 2004 [1, 2], a large number of HEAs and their phases have been reported. The rapidly expanding database, in recent years, made it possible to bring ML into this field [35, 36, 40–44, 54, 87, 115–124]. ML, in general, is capable of extracting non-linear correlation between input and output data. When applied to HEAs it can be utilized to discover patterns in the large amounts of existing HEA data. For ML, each HEA datum includes values of features and a class. The class is the HEA phase. The features are correlated with the phase formation and are utilized to make phase predictions. ML algorithms are methods capable of identifying patterns between the input features and the HEA phases. Based on these connections, phase predictions for new HEAs are given. In ML, choosing informative, discriminating, and independent features is crucial to the training process of an algorithm. The algorithms examine feature-phase relationships in a portion of the whole database called a training set. After that, ML makes and verifies the predictions for HEAs in the remaining database called a test set. A prediction success rate is generated from the ability of the training set to predict the test set correctly. The current HEA ML prediction methods can differ by three different aspects: (1) the features used, (2) the algorithms used to analyze the training database, or (3) the HEA phase classifications.

Tancret et al. [35] combined empirical parameters ($\Delta\chi$, VEC, $K_m$, $\Delta H_{mix}$, $\delta$, $\mu$, $e/a$, $\Omega$, and $S_m$), Gaussian processes (GP) ML algorithm, and CALPHAD to find a robust method of identifying single-phase HEAs in a database with 322 HEAs. The use solely of empirical parameters or CALPAHD is not reliable in phase prediction. Nevertheless, the combination of the two with GP can be useful. GP with empirical features first returns the probability for a HEA being a single-phase SS. When this probability for a HEA was higher than 0.59, CALPHAD would be applied to predict the phase. All HEAs were found to be single-phase SS when both predictions agree. However, many single-phase SS HEAs were misidentified as mixed phase HEAs by this method, this led to the absence of potential useful HEAs from the predictions.

GP was also used by Pei et al. [54] in classifying alloys as a multi-phase or a single-phase. Single-phase alloys were further classified as a BCC, FCC, or HCP phase. The database included 1,252 alloys ranging from binary alloys to multi-component HEAs. The database was partitioned into 627 multi-phase alloys and 625 in the single-phase alloys. The atomic percentage weighted averages of 85 elemental properties composed the features pool. Initially, ten features were selected based on their relevance for making GP phase classification decisions. Then different combinations of features were tested until only the optimum features remained. This method returned a prediction accuracy of 93 %. Molar volume, bulk modulus, electronegativity, melting temperature, valence, vaporization heat, and thermal conductivity were determined to be the most relevant features. GP returned a probability for each alloy indicating its tendency to form a single phase. This GP probability was plotted against δ in Fig. 4, where GP probability > 0.5 and δ < 6 % were the criteria for forming a single phase.

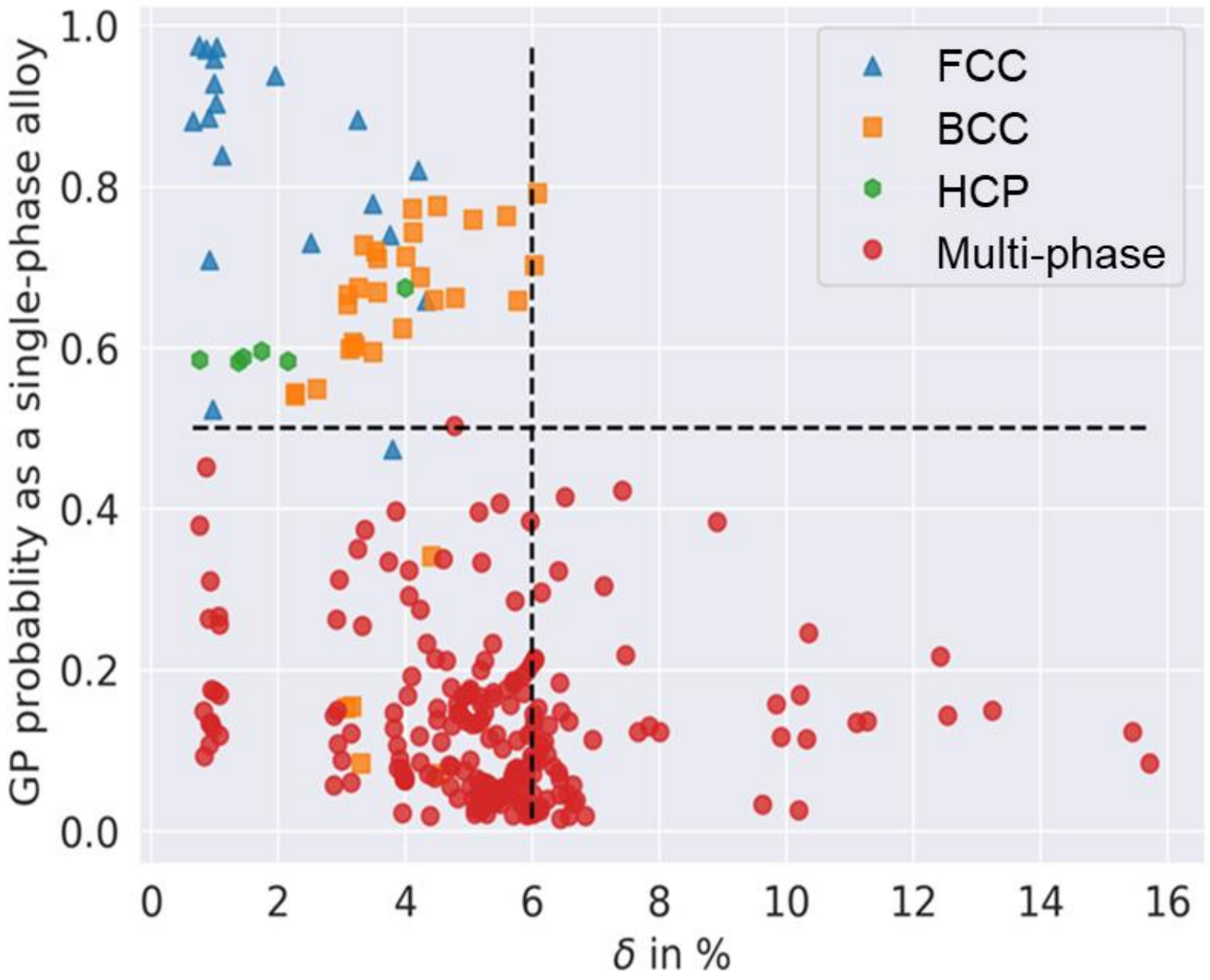

Figure 4. GP probability as a single-phase alloy plotted against radius mismatch, δ to separate alloys into different phases. The criteria for forming a single-phase alloy are represented with dashed lines. Figure from Pei et al [54].

Islam et al. [40] used empirical parameters as the ML features as well, but with an ANN ML algorithm. The 118 HEAs used were classified as 64 SS, 21 IM, and 33 amorphous (AM) phases. ANN predictions showed that the relevance with the phase formation between different features increases in the following order: $\Delta S_{mix}$, δ, $\Delta H_{mix}$, Δχ, and VEC. The ANN prediction accuracy was 83 %. Like prior mentioned methods, detailed phase content is still not predictable with this method.

Huang et al. [41] used five empirical parameters VEC, Δχ, $\Delta H_{mix}$, $\Delta S_{mix}$, and $\delta$ as ML features. The 401 HEAs used were classified as 174 SS, 173 SS+IM, and 54 IM HEAs. Three kinds of ML algorithms were used; (1) k-nearest neighbors (KNN) returned a prediction accuracy no larger than 68.6 %, (2) SVM returned an accuracy of 64.3 %, and (3) supervised multi-layer feed-forward neural network (MLFFNN) returned an accuracy of 74.3 %. Binary classifications of phases between SS and IM, SS and SS+IM, as well as IM and SS+IM, were also conducted with MLFFNN, returning accuracies of 86.7 %, 78.9 %, and 94.3 %, respectively. Of the empirical parameters used, $\delta$ and VEC were of greater importance than the others. According to the authors, additional features will improve the accuracy.

Similarly, Li and Guo [42] used SVM with empirical parameters to classify a database containing 18 BCC, 43 FCC, and 261 other phases called NSP (not forming single SS phase) HEAs. Different combinations of ML features among candidates {VEC, $\delta$, $\Delta H_{mix}$, $\Delta S_{mix}$, $\Delta\chi$, $\Delta H_f$, $T_m$} were used. The feature combination of {VEC, $\delta$, $\Delta H_{mix}$, $\Delta S_{mix}$, $T_m$} was found to give the best prediction accuracy of 90.69 %. The test accuracy increases by > 5 % when the training data set percentage increased from 50 % to 90 %. This indicates that the performance of the model can be further increased by including additional future experimental data. When the training set percentage was 90 %, the accuracies for BCC, FCC, and NSP phase predictions were 60 %, 75 %, and 97.79 %, respectively. The error in the performance occurred due to missed predictions between FCC and NSP, or BCC and NSP phases. The method excelled at separating BCC and FCC HEAs.

Agarwal and Prasada Rao [44] used an adaptive neuro-fuzzy interface system (ANFIS), a hybrid method using an ANN ML algorithm and fuzzy logic, to predict HEAs with BCC, FCC, and multi-phases. Two sets of input features were used. The first used compositions of HEAs and returned a prediction accuracy of 84.21 %. The second used empirical parameters VEC, $\delta$, $\Delta H_{mix}$, $\Delta S_{mix}$, $\phi$, and $\sqrt{<\varepsilon^2>}$ and returned a prediction accuracy of 80 %. In the second model, the importance of each feature was ranked by removing one individual empirical feature and calculating the prediction accuracy drop due to the absence. The ranking of importance of the empirical features was determined to be $\sqrt{<\varepsilon^2>}$ > VEC > $\delta$ > $\phi$ > $\Delta H_{mix}$ = $\Delta S_{mix}$. By systemically changing the value of one feature while keeping the other features unaltered, the phase prediction results from ANFIS could reflect how the change of each feature affects the phase formation. For example, BCC phase formation is favored over FCC phase formation when $\delta$ increases.

Zhou et al. [43] used three algorithms to study the phase formation rules. They were ANN, one-dimensional convolutional neural network (CNN), and SVM. The database used was composed of 13 empirical features and 601 as-cast binary, ternary, quaternary, and higher-order alloys. The ML model studied the appearance of SS, IM, and AM phases in HEAs. Multiple positive phase predictions would indicate a combination of those predicted phases. The testing accuracies of three algorithms on predicting the appearances of the SS, IM, and AM phases were all near or above 95 %. Correlations between features and the appearance of a phase were examined by a compound transformation function. It was derived from linear transformation matrices and biases among the input, hidden, and output layers in the ANN model. As can be seen in Fig. 5, certain features were found to be strongly correlated to specific phase appearances. For example, large values for $T_m$ and the standard deviation of binary $\Delta H_{mix}$ ($\sigma_{\Delta H}$) promote the formation of the AM phase while suppressing the formation of the IM phase. And while $\Delta S_{mix}$ promotes, $\delta$ suppresses the SS phase formation. Experimental results using $(FeCrNi)_{10-x}(ZrCu)_x$ further validated this model, however, the results are cooling rate sensitive.

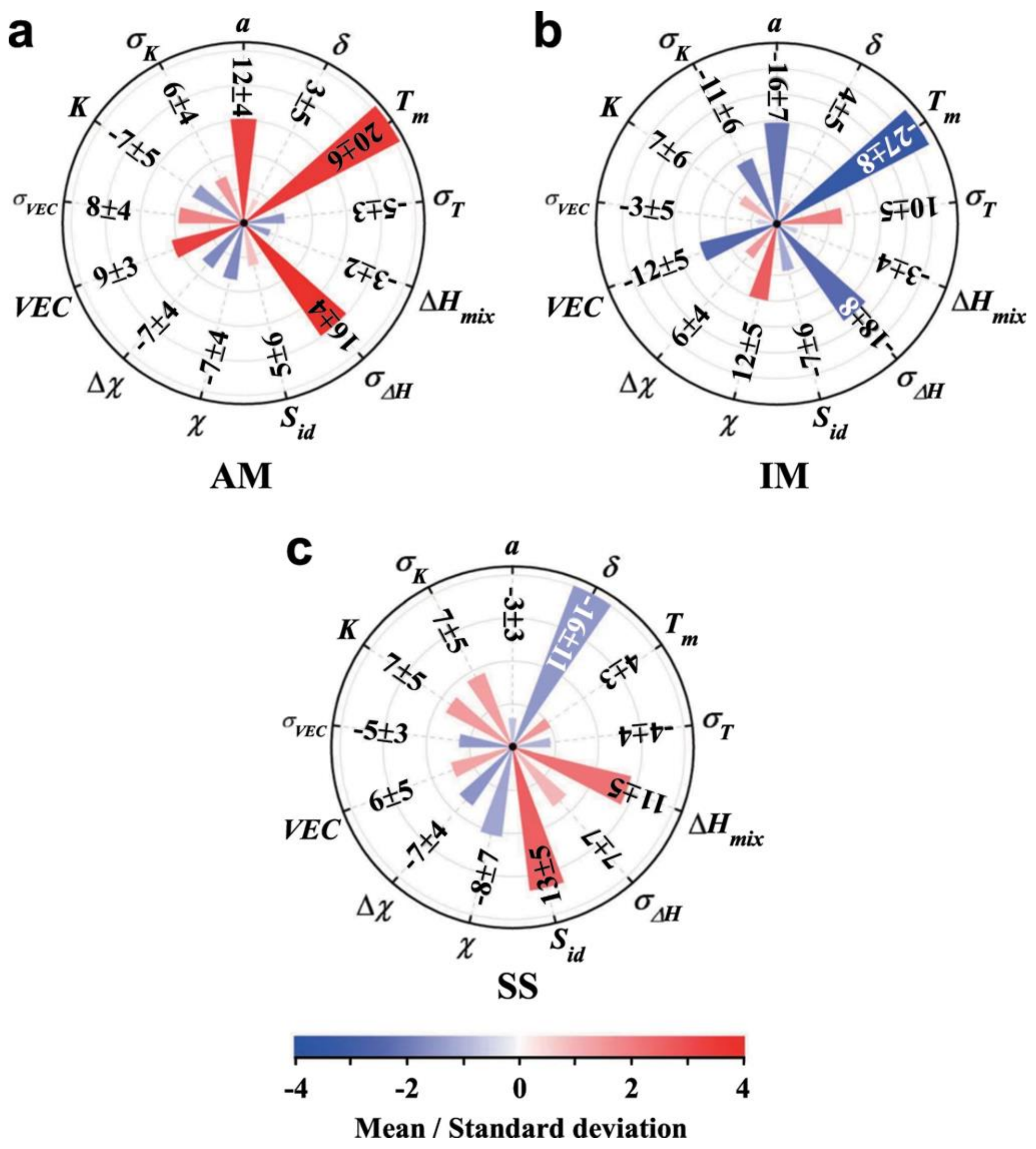

Figure 5. Correlations between features and the appearance of (a) AM, (b) IM, and (c) SS phases. Red and blue colors represent positive and negative correlations, respectively. Figure from Zhou et al [43].

Recently, there has been an increase in studies utilizing feature engineering and active learning [87, 88, 118, 125, 126]. Feature engineering is an approach to mathematically expanding the ML feature pool. A large number of features are synthesized from a limited basis set of features. Then a down-selected combination of features is chosen, which returns the highest accuracy. Active learning is an approach that can experimentally expand the database under the guidance of ML and improve prediction accuracy. They have been applied in the following works.

Zhang et al. [118] utilized the genetic algorithm (GA) method to select the best combinations of ML features and models systemically. In their classification I, a 550 HEA database was classified into SS and non-solid solution (NSS) phases. In their classification II, the SS HEAs were further divided into FCC, BCC, and dual phases (DP), a combination of FCC and BCC phases. A flowchart for this work is shown in Fig. 6. The pool of features and models was composed of 70 features and nine common ML classification algorithms. The results from their computations showed that only a minimum of four features were required to produce an accurate prediction. The Pearson correlation coefficients [127], which measures the statistical relationships between two continuous variables, were calculated among the 70 features to remove the redundant features. For each ML algorithm, the GA systemically changed the four features used and determined the feature combination returning the best prediction accuracy. All nine algorithms were optimized in the same manner, and the best model was selected. Classification I and II eventually had accuracies of 88.7 % and 91.3 %, respectively. Active learning was further employed to refine their predictions. Ten new HEAs, whose predicted phases had high uncertainties from ML, were experimentally prepared and measured for their phases. After adding the new data into the database, the ML prediction accuracy increased. This implies that iterating the active learning steps can improve the ML accuracy.

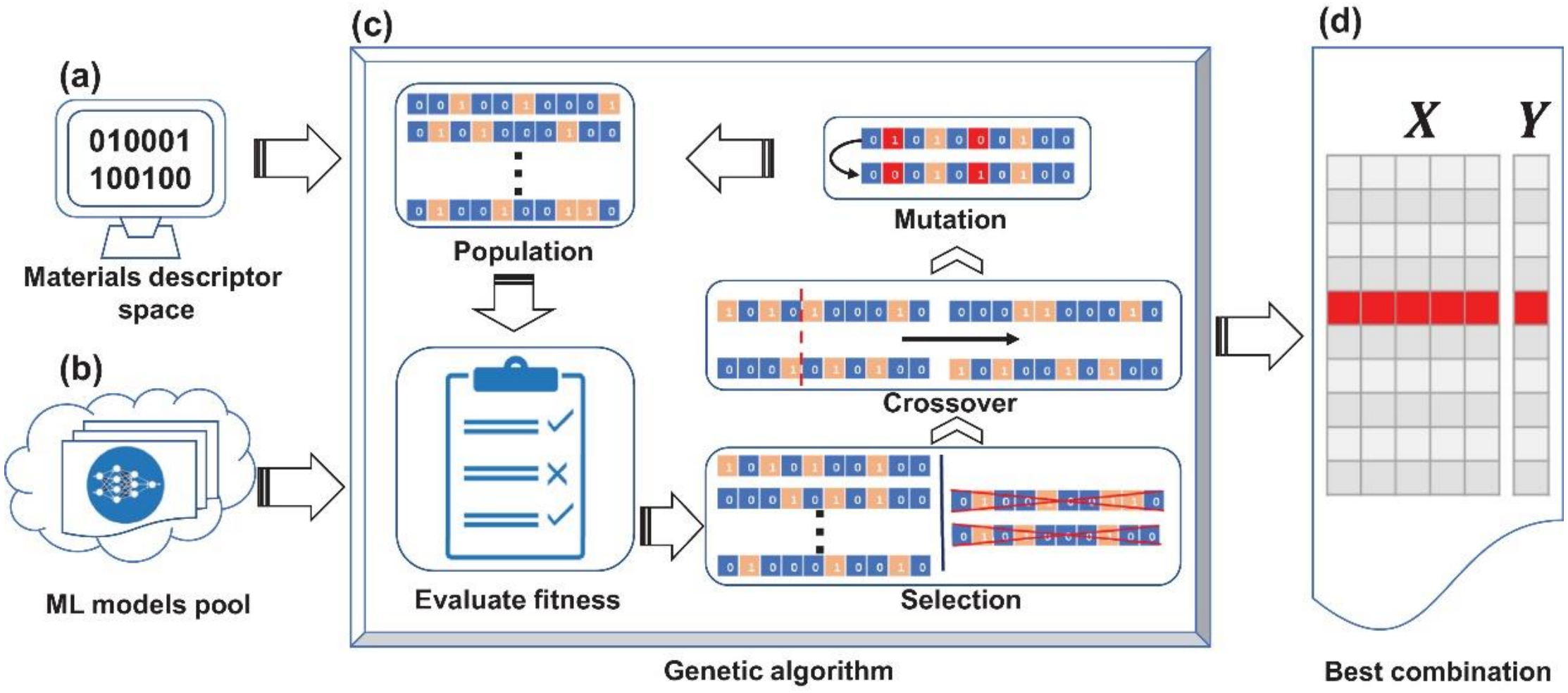

Figure 6. Flowchart from Zhang et al. [118] describing the GA method to best select ML algorithms and features for HEA selection. First, (a) is the ML feature pool with 70 features that can influence the phase formation. Second, (b) is the commonly used ML algorithms pool. Third, (c) is the GA process, where the optima combinations of features and ML algorithm is obtained. And lastly, (d) is the optimized output.

Dai et al. [119] explored feature engineering with empirical features to improve the prediction ability. First, specific features were highly correlated with each other based on their Pearson correlation coefficients. Feature pairs such as $\sqrt{<\varepsilon^2>}$ and $\delta$ had large Pearson correlation coefficients and one redundant feature should be removed. With this method, 14 candidate features were down-selected to nine features. Second, from the nine features, a pool of 30,450 non-linear features was generated. These non-linear features were calculated by the following relationships: $\sqrt{|x|}$, $x^2$, $x^3$, and log(1+|x|) for feature x, or multiplying two or three of these features together. A recursive feature elimination method was used to eliminate the irrelevant features, and 20 features were left. To compare the two feature pools, a database with 407 HEAs and a simple linear regression ML algorithm were employed to classify the HEAs into BCC, FCC, HCP, multi-phase, and AM phase categories. The highest ML prediction accuracies obtained from using the original nine features and the 20 engineered features were 75 % and 86 %. The constructed non-linear features outperformed the original features. This work shows that feature engineering can improve the feature-phase relationship to increase the accuracy of predictions.

Empirical parameters are frequently selected as features for ML methods. The following ML methods limit features to only the HEA compositions or their elemental components.

Wu et al. [120] designed a eutectic HEA system, AlCoCrFeNi, using ANN. Their database contained 311 eutectic HEAs predicted by CALPHAD and 10 HEAs discovered by experiments. Their data was divided into training and test sets of 75 % and 25 %, respectively. The input features consisted of the atomic percentages of the five elements in each HEA. The output was a normalized number between -1 and +1: a negative value represented the formation of a hypereutectic B2 primary HEA, 0 was the formation of an eutectic HEA, and a positive value represented the formation of a hypoeutectic FCC HEA. As shown in Fig. 7a, the predicted values and the target values are in agreement. The ML showed high phase constitution prediction ability, and 400 new near-eutectic HEAs were predicted. Fig. 7b-c are the element content distributions of different elements for the new near-eutectic HEAs. In Fig. 7b, the Al and Cr content distributions are clustered in certain ranges, while the Co, Ni, and Fe content distributions are near evenly distributed. Fig. 7c shows correlations between different element content distributions. A strong content correlation for Al-Cr is noted. Additional findings also seen from Fig. 7: Al was identified as the most relevant element to determine the phase constitution; Cr is associated with Al content to influence the eutectic formation; Co, Ni, and Fe are miscible elements influencing the eutectic formation by tuning their average VEC value; and a high VEC favors a FCC phase formation while a low VEC favors a B2 phase formation. From their work, an effective eutectic HEA design pathway was presented, and several HEAs exhibiting strengths of ~1300 MPa and elongation of ~20 % were made.

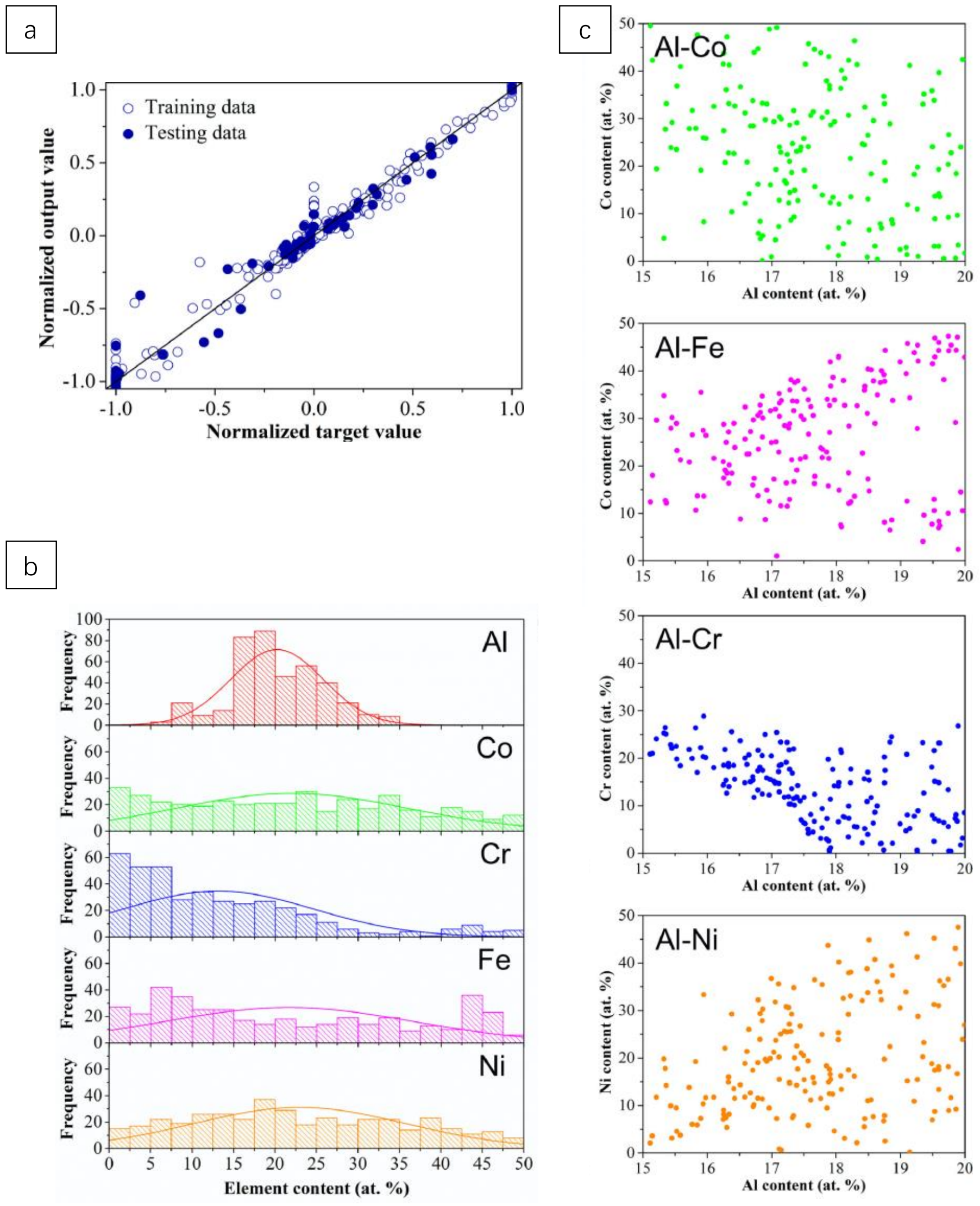

Figure 7. (a) Regression analysis of the training and testing data. The predicted values and the target (actual phase constitution) values are the y and x-axes. (b)-(c) Element content distributions of different elements for the 400 predicted near-eutectic HEAs. Figures from Wu et al [120].

Kube et al. [36] used a linear ordinal logistic regression method to predict the HEA phases based on their compositions. After optimization, the values were assigned to elements representing their strength in stabilizing BCC or FCC phases. The average of these stabilizing effects, denoted as solid solution selection index (SSSI), determined HEA phase formation tendencies. As shown in Fig. 8, this method separates the BCC and FCC HEAs, but not the mixed BCC+FCC HEAs. The results showed that certain elements have an influence on stabilizing specific phases. There are three particular limitations to this method. First, the HEA database used for training sets was produced from high-throughput sputter depositions, which as a method can extend the compositional range of SS phase formation due to the rapid quench nature of sputtering as opposed to phase formation from traditional alloying methods. Second, the elements involved in this study were confined to Al, Cr, Mn, Fe, Co, Ni, and Cu with the effect of other elements on phase formation needing further study. And final limitation is that the phases predicted by this method are only BCC, FCC, and their mixtures while other phases are not accounted for.

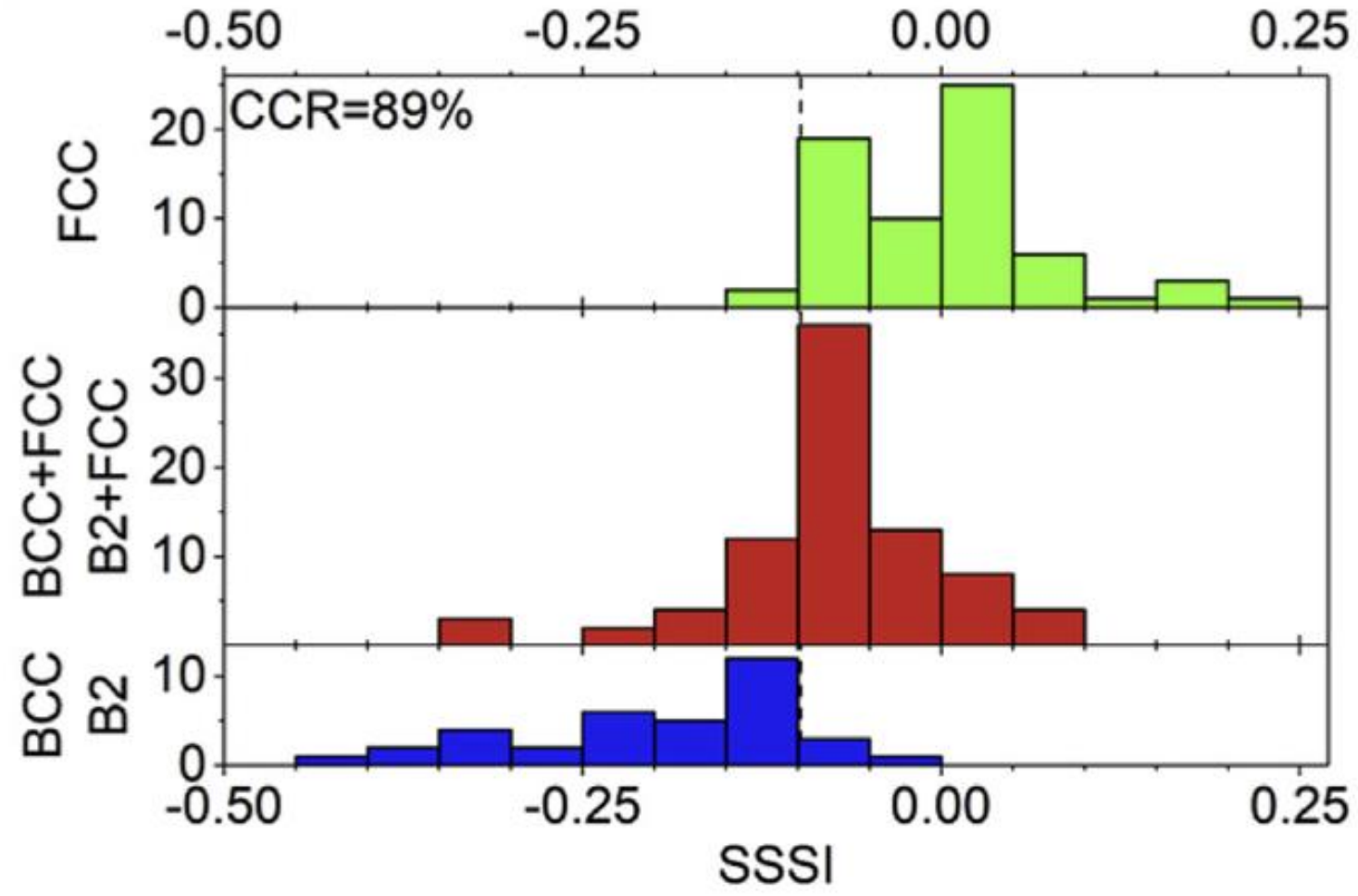

Figure 8. Prediction performance of Kube's method. SSSI is the parameter representing a phase stabilizing effect. Negative SSSI favors BCC/B2 phase formation, while positive SSSI favors a FCC phase formation. Blue (bottom), red (middle), and green (top) histogram bar plots correspond to BCC/B2, BCC+FCC/B2+FCC, and FCC phases, respectively. BCC/B2 and FCC HEAs are mostly separated by their SSSI values, while the mixed BCC/FCC phases still mix with other phases. The overall correct classification rate (CCR) is 89 %. Figure from Kube et al [36].

The prior mentioned statistical and ML methods are summarized in Table.1. For ML, the empirical parameters are the most frequently used features. ML combinations of these features yield an improved robust prediction ability as opposed to any of the features used individually. Compared to the first-principles calculations, ML is not computationally intensive and is not limited by the availability of thermodynamic databases. CALPHAD, however, is superior at predicting detailed phase compositions. The limit of the ML methods results from database limited availability of less common phases. The prior summarized ML methods can predict HEA broad phase categories such as SS+IM, but they struggle with specific phase categories like BCC+B2.

Table 1. Summary of the current statistical and ML methods for predicting HEA phases. Abbreviations of the algorithms are: ANFIS (Adaptive Neuro-Fuzzy Interface System), ANN (Artificial Neural Network), CALPHAD (Calculation of Phase Diagrams), CNN (Convolutional Neural Network), GA (Genetic Algorithm), GP (Gaussian Processes), KNN (K-nearest Neighbors), MLFFNN (Supervised Multi-layer Feed-forward Neural Network), and SVM (Support Vector Machine). The ML classification algorithms, the ML features, the phase categories defined in each ML method, the prediction success rates, and the references to the work are listed. In the phase categories column, the total count of HEAs in each phase category is listed, if the information was available.

| Classification Algorithms | Features | Phase Categories (count reported) | Overall Success Rate | Ref. |
|---|---|---|---|---|
| ANFIS | HEA Compositions | HEA: BCC, FCC, and | 84.21 % | Agarwal and |
| ANFIS | VEC, $\delta$, $\Delta H_{mix}$, $\Delta S_{mix}$, $\phi$, and $\sqrt{<\varepsilon^2>}$ | Multi-phase | 80 % | Prasada Rao[44] |
| ANN | HEA Compositions | HEA (321): Hypereutectic, Eutectic, and Hypoeutectic | N/A | Wu et al.[120] |
| ANN | $\Delta S_{mix}$, $\delta$, $\Delta H_{mix}$, $\Delta\chi$, and VEC | HEA: SS (64), IM (21), and AM (33) | 83 % | Islam et al.[40] |
| ANN, CNN, and SVM | 13 Empirical Parameters | HEA (601): SS, IM, and AM | > 95 % | Zhou et al.[43] |
| Feature Engineering + Simple Linear Regression | 20 Features engineered from 14 Empirical Parameters | HEA: BCC (43), FCC (48), HCP (16), Multi-phase (237), and AM (63) | 86 % | Dai et al.[119] |
| GA + Active Learning | 4 Features downselected from 70 Features by Feature Engineering | HEA (550): SS, and NSS | 88.7 % | Zhang et al.[118] |
| | | SS HEAs: FCC, BCC, and DP | 91.3 % | |
| GP + CALPHAD | $\Delta\chi$, VEC, $K_m$, $\Delta H_{mix}$, $\delta$, $\mu$, $e/a$, $\Omega$, and $S_m$ | HEA (322): Single SS, and Other phases | 63 % to 80 % (single SS, CALPHAD database dependent) | Tancret et al.[35] |
| GP | Atomic Percentage Weighted Averages of 85 Elemental Properties | HEA & Non-HEA (1,252): BCC, FCC, HCP, and Multi-phase | 93 % | Pei et al.[54] |
| KNN, MLFFNN, SVM | $\Delta S_{mix}$, $\delta$, $\Delta H_{mix}$, $\Delta\chi$, and VEC | HEA: SS (174), SS+IM (173), and IM (54) | 74.3 % (MLFFNN), 68.6 % (KNN), 64.3 % (SVM) | Huang et al.[41] |
| SVM | $\Delta S_{mix}$, $\delta$, $\Delta H_{mix}$, $T_m$, and VEC | HEA: BCC (18), FCC (43), and non-single-phase (261) | 60 % (BCC), 75 % (FCC), 97.79 (NSP) | Li and Guo [42] |
| Linear Ordinal Logistic Regression | HEA Compositions (generated from sputtering deposition) | HEA: BCC or B2 (762), FCC (553), and Mixed BCC or B2 + FCC (446) | 89 % | Kube et al.[36] |

**Section 3:** Phenomenological Approach

### 3.1 Motivation

Experimental phase measurements can differ from theoretical predictions. Factors unaccounted for could influence the formation of these phases. To include theses unknown factors into the prediction of phases, a phenomenological predictive method [45] based on experimental outcomes is required. Experimentally determined binary phase diagrams contain information on the crystal structure, elemental mixing, and phase separation over temperature and composition ranges. They have encoded within them the information for equilibrium binary phase formation. Information for the prediction of the phase of a HEA can be extrapolated from the set of all possible constituent binary phase diagrams. Here, one such phenomenological predictive method devised by the authors of this review article is presented.

First, the method defines several phenomenological parameters calculated from binary phase diagrams that influence HEA phase formation tendencies. Next, a database covering the majority of known experimentally validated HEAs was processed using a ML data mining technique. Finally, to verify the effectiveness of this method, the results were tested through experimental techniques.

The HEA phase prediction method discussed in this section represents a subset within a larger machine learning model which is developed to design HEAs with desirable structural and functional properties. To provide a high-level view of the method, a flowchart illustrating the flow of processes within the ML model is shown in Fig. 9. The use of phenomenological features for predicting HEA phases with certain homogeneity ranges is the focus of this section. Illustrative examples of employing adaptive features to predict intermetallic phases are discussed in Section 5.

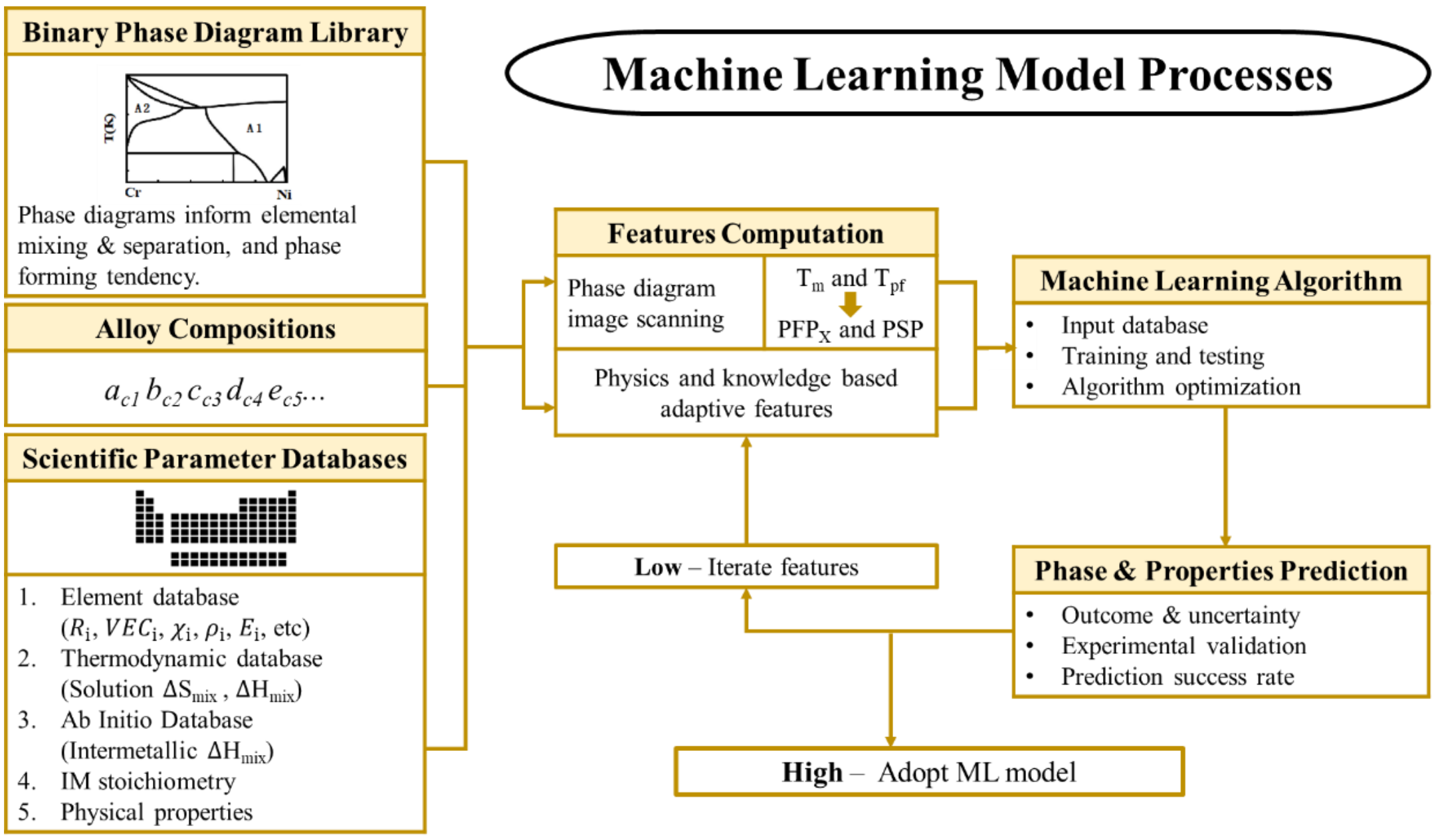

Figure 9. Flowchart illustrates machine learning model for predicting HEA phases. The acronyms and symbols are defined in Section 2 and this Section 3.

The work presented in this chapter has been updated from the originally reported results [45], with additional reported HEAs added into the ML database. Currently, there are over 1,100 reported HEAs in the updated version of the prior reported comprehensive database [45]. Only a subset of 828 of the HEAs was used. They fulfilled two requirements: (1) they were in either as-cast or high-temperature annealed states, and (2) they only have phases with broad homogeneity regions, e.g., FCC, BCC, HCP, B2, Sigma, or Laves phases. HEAs were classified based on their phase content into six categories: disordered FCC (**A1**), disordered BCC (**A2**), disordered HCP (**A3**), mixed disordered FCC+BCC (**A1+A2**), B2 mixed with other disordered SS phases A1, A2, or A3 (**B2+SS**), and either Sigma or Laves IM mixed with other phases (**IM+**).

## 3.2 Description of the method

### 3.2.1 Parameters from binary phase diagrams

Here several parameters related to phase formation in HEAs are defined. There are two types of parameters involved in this method: (1) phase field, and (2) phase separation. These parameters were dependent on one of two temperatures where phase transformation was arrested.

One HEA phase-locking temperature was the undercooling temperature near the $T_m$ and the other was an alloy postproduction annealing temperature. For this section, the melting temperature of the HEA must be defined. $T_m$ here was determined from the binary phase diagrams by using a weighted average of the binary melting temperature as follows:

$$T_m = \frac{\sum_{i \neq j} T_{i-j} \times c_i \times c_j}{\sum_{i \neq j} c_i \times c_j} \quad (\text{Eqn. } 18)$$

where $T_{i-j}$ is the melting temperature of the i-j elements for the relative ratio of the two elements from the HEA composition. For the as-cast HEAs, the undercooling temperature extends to 0.8 $T_m$. The phase formation temperature ($T_{pf}$) was introduced and defined as the temperature where rapid phase evolution ceases and the phases formed are retained post quench. $T_{pf}$ was approximated as $T_{pf}$ = 0.8 $T_m$, which optimized the results for the ML predictions [45] from this work. For the annealed HEAs, the phases formed during annealing are locked in with rapid quenching, and $T_{pf}$ is assigned the final annealed temperature. The phenomenological parameters that controlled the phase formation were calculated based on the $T_{pf}$ value.

1. Phase Field Parameter:

When the temperature is above or equal to $T_{pf}$ atoms are free to exchange neighbors due to high atomic mobility. The neighbors of each atom are random. The alloy mixture is essentially ergodic and local atoms have nearly equal probabilities of sampling any binary configurations on the relevant phase diagrams. As such, the probability of forming a phase X locally for i-j elements can be determined by the binary phase field percentage of phase X on an i-j phase diagram and is denoted as $X_{i-j}$.

The local probabilities of forming a specific phase from all atomic pairs can be integrated to yield an overall probability. The probability of forming a phase X for the HEA is the Phase Field Parameter ($PFP_X$), and it is calculated as the weighted average of all constituent $X_{i-j}$ by Eqn. 19.

$$PFP_X = \frac{\sum_{i \neq j} X_{i-j} \times c_i \times c_j}{\sum_{i \neq j} c_i \times c_j} \div 100\ \% \quad (\text{Eqn. } 19)$$

In this method, the $PFP_X$ values have been calculated for the targeted phases, and they are denoted as $PFP_{A1}$, $PFP_{A2}$, $PFP_{B2}$, $PFP_{A3}$, $PFP_{Laves}$, and $PFP_{Sigma}$.

HEA $Al_2CoCrCuNi$ is presented as an example to determine the phase field percentages used to calculate $PFP_X$. This HEA has a predicted $T_m$ = 1569 K and the phases are assumed to be locked at $T_{pf}$ = 1255 K. In Fig. 10, it is seen that high concentrations of Cr favor BCC phase formation, while high concentrations of Ni favor FCC phase formation. Under the assumption of equally sampling all binary configurations, the probability of Cr-Ni binary favoring BCC phase formation locally is the binary phase field percentage of the BCC phase. This percentage is the line segment between the two intersection points of an isotherm at $T_{pf}$ and the compositional boundary of the BCC phase. In this case, it is approximately 5 % for the BCC phase and approximately 44 % for the FCC phase.

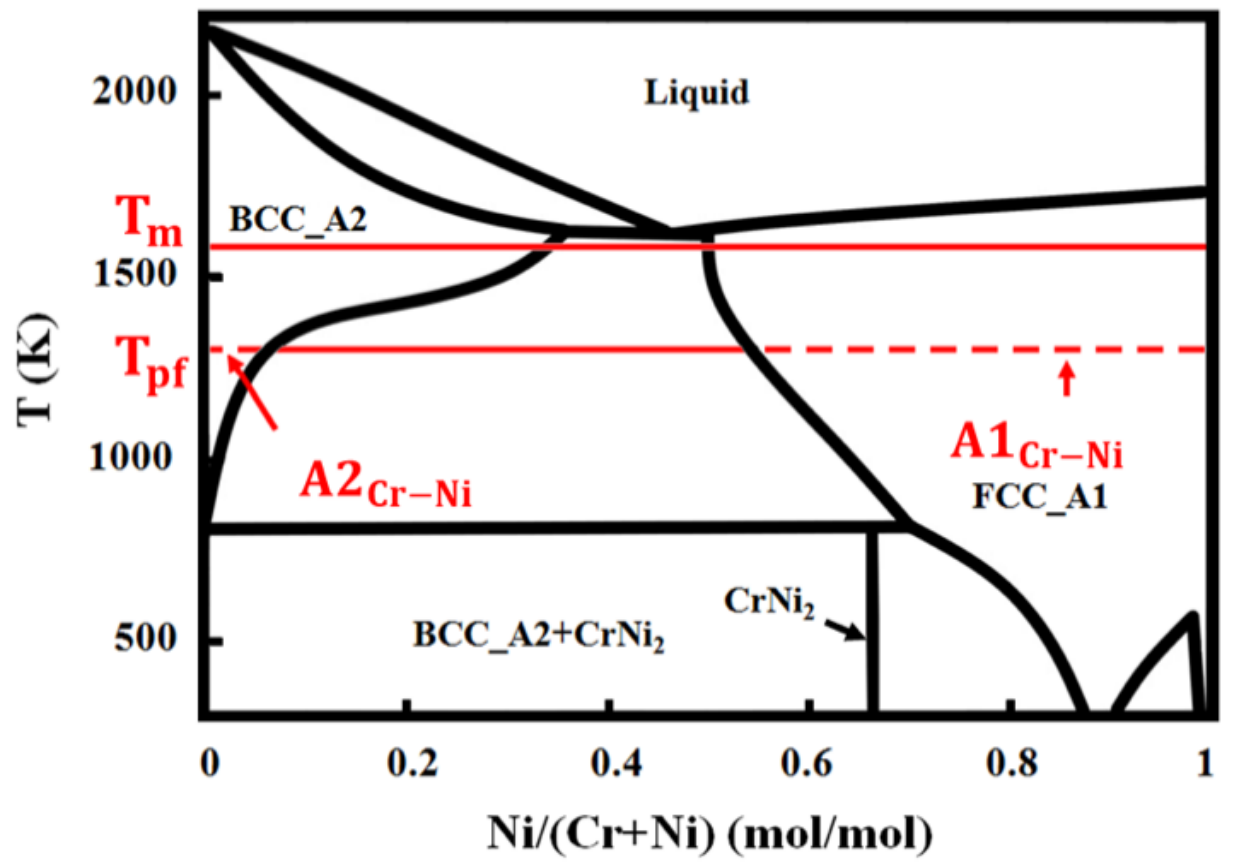

Figure 10. Demonstration of the binary phase field percentage calculation. The binary phase diagram Cr-Ni is used to determine the fractions of BCC and FCC phases for the HEA $Al_2CoCrCuNi$. The binary phase field percentages of BCC and FCC phases are represented as $A2_{Cr-Ni}$ and $A1_{Cr-Ni}$, respectively. Figure from Qi et al [45].

These binary phase field percentages are then used to calculate $PFP_X$, which will be used to visualize separations in the HEA phase space. The above calculation method can be applied to any phase diagram and for any type of phase.

2. Phase Separation Parameter:

If a miscibility gap exists in phase diagrams, this interatomic repulsion can lead to phase separation in HEAs [128, 129] and the formation of multiple coexisting phases such as FCC+BCC. The binary phase separation percentage on the binary phase diagram represents the probability of the two elements being separated into two different phases in the HEA. For a given phase diagram, an isothermal line drawn at $T_{pf}$ is composed of two parts. The first is the binary phase separation percentage denoted as $\mathrm{Separation}_{i-j}$ and the remainder of the line is defined as the elemental mixing denoted as $\mathrm{Mixing}_{i-j}$ for an i-j binary system. If the phase separation is absent from a phase diagram, then $\mathrm{Separation}_{i-j} = 0$ %. To calculate the Phase Separation Parameter (PSP) for a HEA the following equation is used

$$\mathrm{PSP} = \frac{\sum_{i \neq j} \mathrm{Separation}_{i-j} \times c_i \times c_j}{\sum_{i \neq j} \mathrm{Mixing}_{i-j} \times c_i \times c_j} \quad (\mathrm{Eqn.}\ 20)$$

where the $\mathrm{Separation}_{i-j}$ and $\mathrm{Mixing}_{i-j}$ are used from the HEA constituent binary systems.

$\mathrm{Separation}_{i-j}$ and $\mathrm{Mixing}_{i-j}$ are illustrated using the same HEA as used to calculate binary phases field for $PFP_X$. Fig. 11 shows two binary phase diagrams of the $Al_2CoCrCuNi$ HEA with different separation effects. In Fig. 11a, the large positive $\Delta H_{mix}$ of the Cr-Cu binary prevents them from having a mixing effect. In HEAs, Cu and Cr tend to reside in the different phases. In Fig. 11b, a large separation effect exists for the Co-Cu binary due to the positive $\Delta H_{mix}$ with a small mixing effect occurring at high temperatures. The Cr-Cu binary phase diagram at $T_{pf}$ has a $\text{Separation}_{\text{Cr-Cu}} = 100\ \%$ and $\text{Mixing}_{\text{Cr-Cu}} = 0\ \%$, and in the Co-Cu binary phase diagram $\text{Separation}_{\text{Co-Cu}} = 92\ \%$ and $\text{Mixing}_{\text{Co-Cu}} = 8\ \%$.

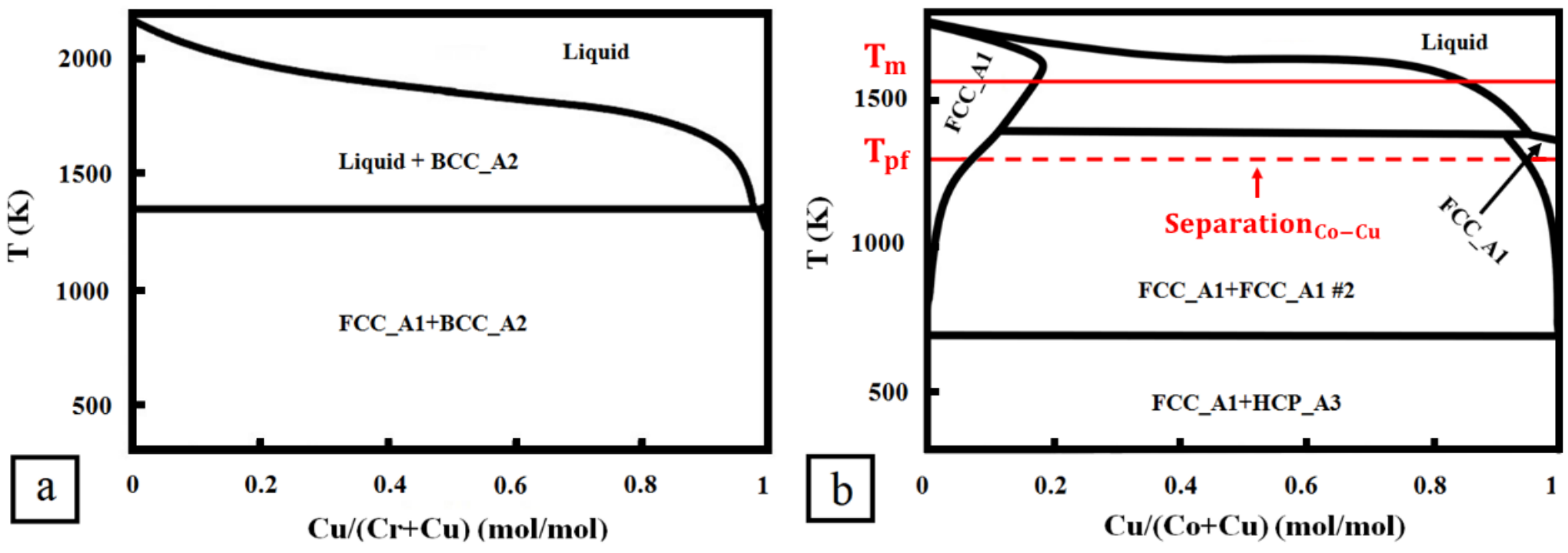

Figure 11. Two binary phase diagrams used to determine the binary phase separation percentage for HEA $Al_2CoCrCuNi$. (a) Phase diagram of Cr-Cu to show a complete phase separation effect. (b) Overlay of the Co-Cu phase diagram illustrating the line segment method to determine the $\text{Separation}_{\text{Co-Cu}}$ for the HEA $Al_2CoCrCuNi$. Figure from Qi et al [45].

### 3.2.2 Phase fields visualization in the feature space

$PFP_X$ parameters were calculated for six phases A1, A2, A3, B2, Sigma, and Laves. The addition of PSP parameter makes seven parameters in total: $PFP_{A1}$, $PFP_{A2}$, $PFP_{B2}$, $PFP_{A3}$, $PFP_{Laves}$, $PFP_{Sigma}$, and PSP. A 7-dimension space with parameter axes was constructed to visualize the distribution of HEA phases. For the 828 alloys studied, these parameters were calculated. To visualize the position of a HEA in this 7D space, several projections in 2D and 3D space were selected. These plots show the partitioning of phase regions for two results based on (1) SS phases and (2) IM phases.

1. SS Phases

The HEA SS phases of A1, A2, A1+A2, B2+SS, and A3 were plotted for various combinations of the phase parameters. Fig. 12 shows different plotted views highlighting HEA phase region separations. These views were selected based on the three parameters which best highlighted distinct HEA phase region separations. Fig. 12a is a combination of the A1, A2, A1+A2, and B2+SS HEA phase regions. In Fig. 12b, A1 and A2 HEAs are separated into high $PFP_{A1}$ or $PFP_{A2}$ regions. A high $PFP_{A1}$ or $PFP_{A2}$ value stabilizes A1 or A2 phase formation, respectively. The A1+A2 HEAs, in Fig. 12c-d, are mostly in a region where neither $PFP_{A1}$ nor $PFP_{A2}$ is dominant. In Fig. 12d, the higher PSP values for A1+A2 HEAs result in separation from the A2 HEAs. PSP prompts the formation of multiple phases due to the elemental repulsion. In Fig. 12e-g, the phase regions of B2+SS HEAs are plotted against phase regions of A1, A2, and A1+A2 HEAs, respectively. $PFP_{B2}$ is used to predict B2 formation. B2+SS HEAs are all located in a region with relatively higher $PFP_{B2}$ values. This indicates that having a high $PFP_{B2}$ value corresponds to having a high probability of forming the B2 phase in a HEA. Fig. 12h shows that all A3 HEAs are separated from the other phases because of high $PFP_{A3}$ values indicating a higher chance of forming A3.

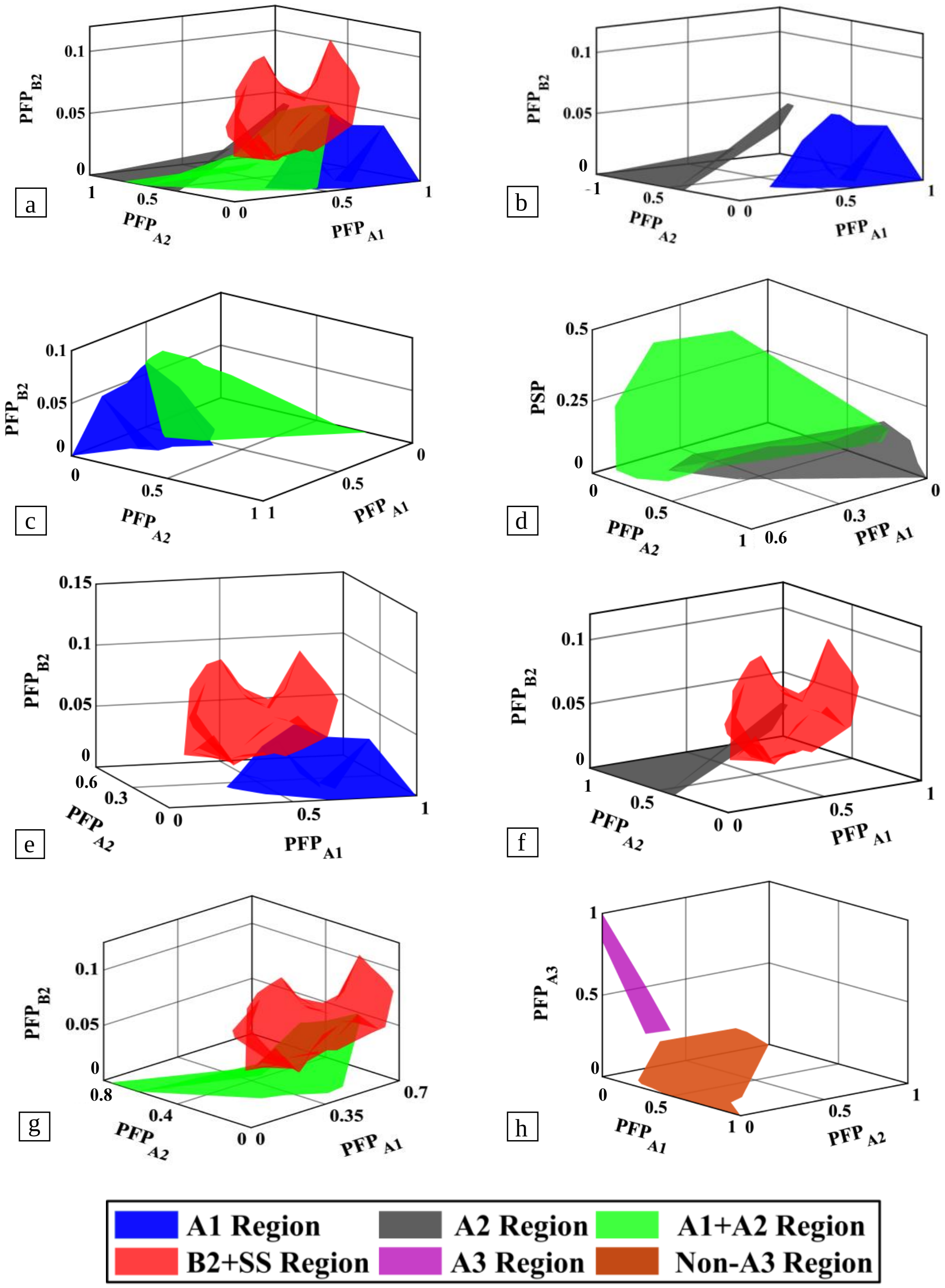

Figure 12. Visualizations of partitions among phases A1, A2, and A1+A2, B2+SS, and A3. (a) $PFP_{A1}$, $PFP_{A2}$, and $PFP_{B2}$ are plotted for A1, A2, A1+A2, and B2+SS HEAs; (b) $PFP_{A1}$, $PFP_{A2}$, and $PFP_{B2}$ are plotted for A1 and A2 HEAs; (c) $PFP_{A1}$, $PFP_{A2}$, and $PFP_{B2}$ are plotted for phase regions of A1 and A1+A2 HEAs; (d) $PFP_{A1}$, $PFP_{A2}$, and PSP are plotted for phase regions of A2 and A1+A2 HEAs; (e)-(h) $PFP_{A1}$, $PFP_{A2}$, and $PFP_{B2}$ are plotted to highlight the B2+SS phase region relative to the A1, A2, and A1+A2 phase regions; and (h) $PFP_{A1}$, $PFP_{A2}$, and $PFP_{A3}$ are plotted for phase regions of A3 and Non-A3 (A1, A2, A1+A2, B2+SS, and IM+) HEAs.

2. IM Phases

Sigma and Laves phases are the two predominant intermetallic phases present in HEAs, based on intermetallic phases present in the HEA database. In Fig. 13, HEAs without IM phase formation (Non-IM) HEAs were plotted with IM+ HEAs on a plot with axes $PFP_{Sigma}$ and $PFP_{Laves}$. Although there is overlap between the phase distribution regions of the IM+ and Non-IM HEAs, IM+ HEAs exist in a region with large $PFP_{Sigma}$ or $PFP_{Laves}$ values.

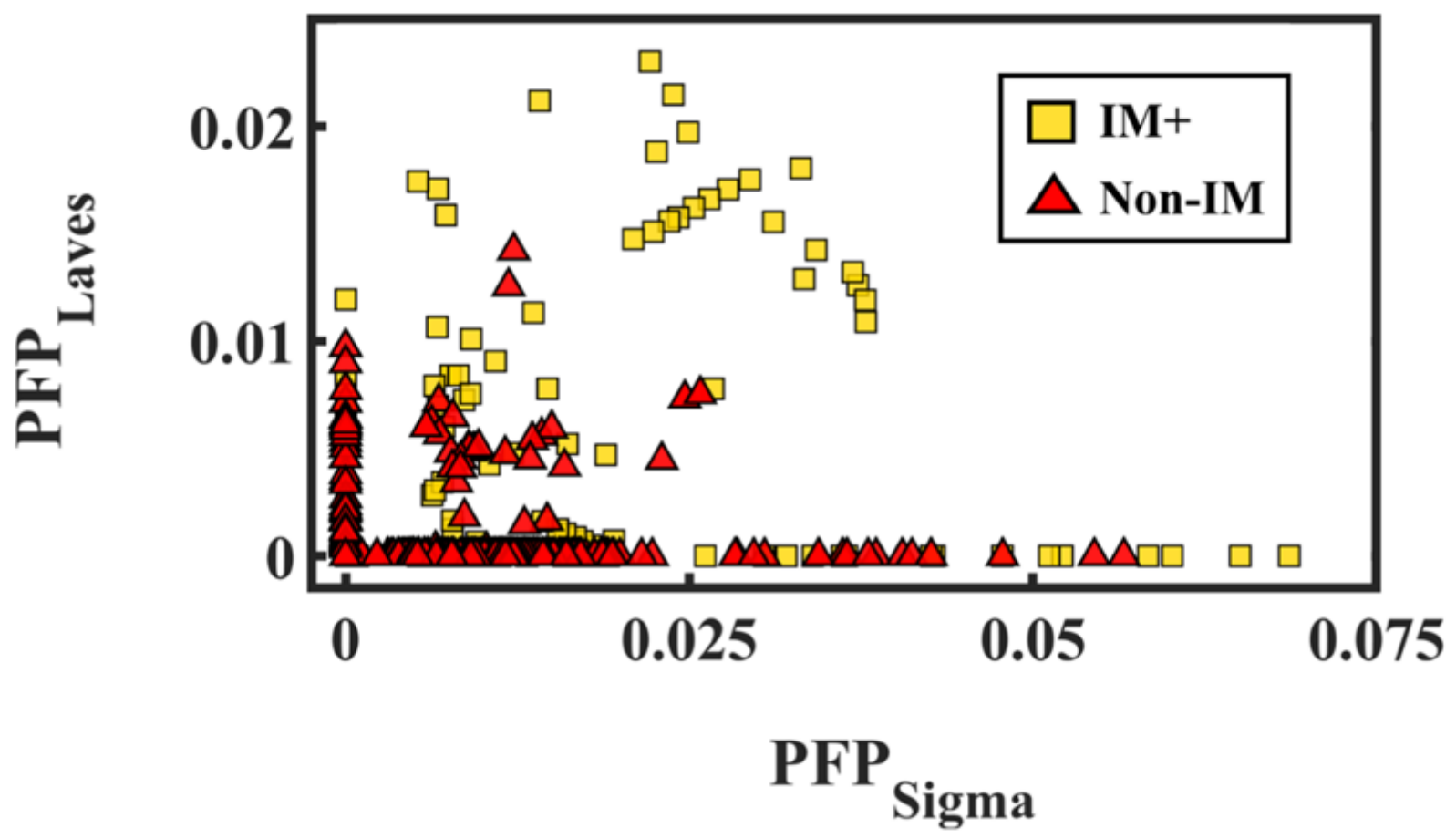

Figure 13. Parameters $PFP_{Sigma}$ and $PFP_{Laves}$ plotted for IM+ and Non-IM HEAs, where Non-IM includes A1, A2, A3, A1+A2, and B2+SS. Figure from Qi et al [45].

The visualization methods for IM and SS HEA phases show that a large $PFP_X$ value generally coincides with the formation of phase X. A large PSP value corresponds with phase separation and multiple phase formations. Different phase regions have overlaps on these plots. Due to the inherent limitations of visualizing seven parameters in 3D space, a better method was needed. Next, a ML method that can include all parameters for a phase formation determination is presented.

### 3.2.3 Optimization by machine learning

ML is the key to solving the visualization limitation issue. The prior defined seven phenomenological phase-diagram based parameters were fed into a ML method as features and they were jointly used to make phase predictions. The ML algorithm called Random Forest was used. The HEA database used was divided into training and test sets, with training set percentages ranging from 10 % to 90 %. Test sets were composed of the remainder of the database.

The phase prediction success rates are shown in Fig. 14. The HEA phase categories are A1, A2, A3, A1+A2, B2+SS, and IM+. With the training set percentage being 90 %, the overall prediction success rate is 83 %. The performance of this method was concluded by the following points:

1. The prediction accuracy was generally higher for the single-phase A1, A2, and A3 and the ordered B2 phase HEAs. The prediction accuracy of the A1+A2 mixed phase is not high. The accurate prediction of the B2 phase was important since the B2 phase has been shown to improve HEA mechanical properties [130].
2. The features are closely correlated to the phase formation. The prediction accuracy decreases only slightly when the training set percentage decreases from 90 % to 50 %. The prediction is accurate even with a small training set. Including new HEAs will only marginally increase the accuracy for these features.
3. When the training set percentage is below 50 %, the success rates drop due to the small training dataset size.

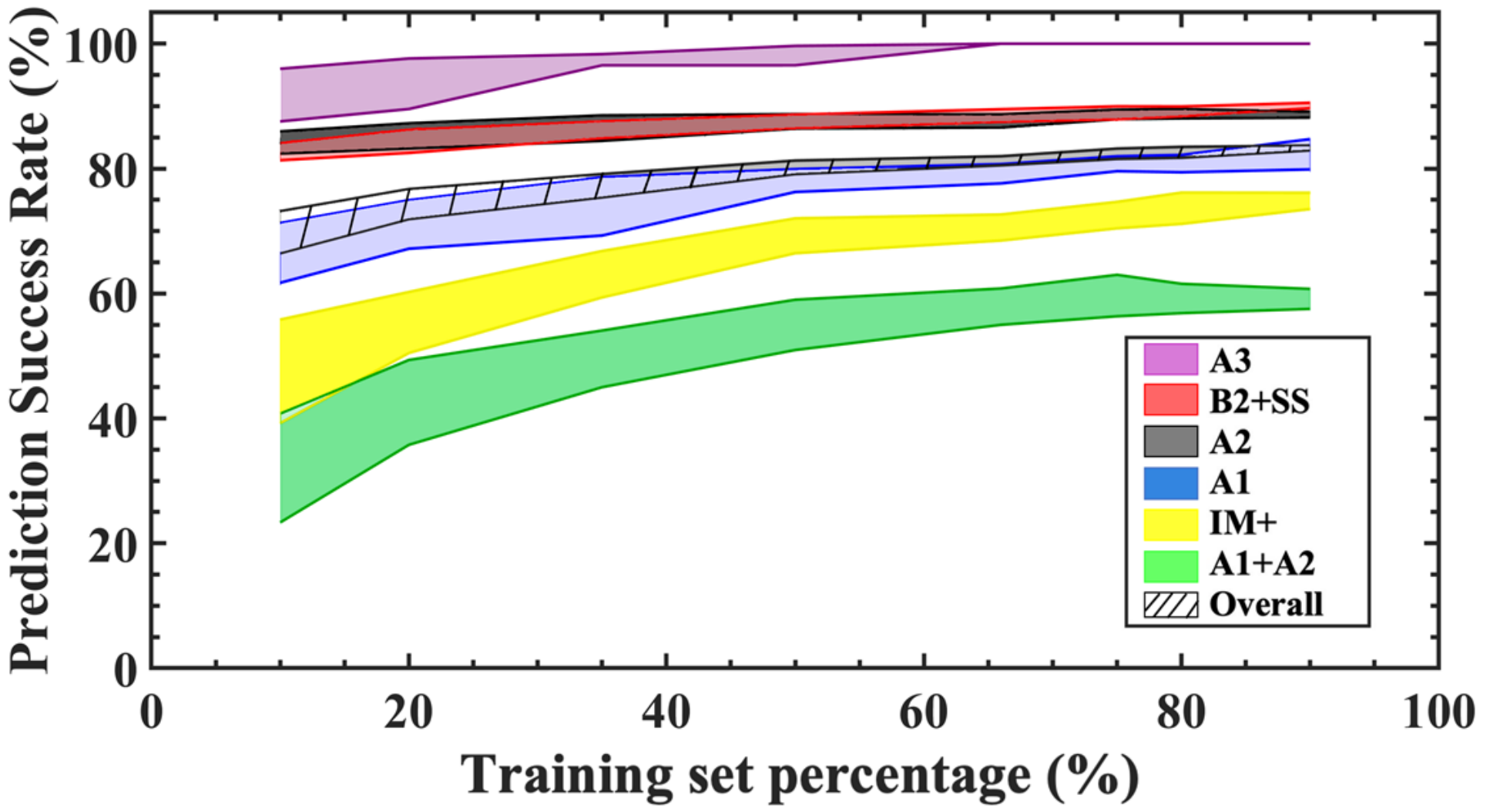

| **Phases** | A1 | A2 | A3 | A1+A2 | B2+SS | IM+ | Overall |
|---|---|---|---|---|---|---|---|
| **Count** | 126 | 178 | 14 | 72 | 290 | 148 | 828 |

Figure 14. ML prediction success rates for HEAs in different phases are plotted. The shaded regions are the confidence bands for prediction success rates for different training set percentages. The width of the confidence bands represents one sigma deviation from the average prediction success rate. In the table are the total counts of as-cast and annealed HEAs for each phase and the overall total HEAs used for the training and test sets.

### 3.2.4 Validation and future development

Validation of this ML method was done by experimentally producing new HEAs. This was done to show that the ML method was not overfitting the database. Forty-four new HEAs with random compositions, listed in Table 2, were synthesized in the as-cast state. Thirty-six of the HEAs were predicted correctly. The accuracy of the prediction was 82 %.

Table 2: HEAs synthesized to validate the ML method. The compositions, predicted phases by the ML, and the XRD measured phases are listed. In the real phase column, the detailed phase information is listed. The eight HEAs whose measured phases differ from predictions are underlined. Modified from Qi et al. [45]

| **Alloy (at. %)** | **Predicted Phase** | **Real Phase** | **Alloy (at. %)** | **Predicted Phase** | **Real Phase** |
|---|---|---|---|---|---|
| $Ag_{0.2}Al_2CrMnNi$ | A1+A2 | **<u>B2, A1, A2</u>** | $CoCrCu_{0.5}FeNi_2Ti_{0.5}V_{0.5}$ | A1 | A1 |
| AgAlCrMnNi | B2+SS | B2, A1, A2 | CoCrCuFe | A1+A2 | A1, A2 |
| $Al_{0.2}CoCr_{0.5}Fe_2NiTi_{0.25}$ | A1 | A1 | $CoCrCuFeMnNiTi_{0.4}$ | A1+A2 | **<u>A1</u>** |
| $Al_{0.2}Cr_{1.5}Cu_{1.5}Fe_{0.5}Mn$ | A1+A2 | A1, A2 | $CoCrCuMn_{0.8}Ti$ | IM+ | Laves, A1 |
| $Al_{0.3}Cr_2Fe_{0.5}Mn_{0.8}$ | A2 | A2 | $CoCrFeMnNi_2V_{0.5}$ | A1 | A1 |
| $Al_{0.5}CoCr_{0.5}CuMnNi$ | A1+A2 | **A1** | $CoCrFeMoNiV_{0.5}$ | IM+ | Sigma, A1 |
| $Al_{0.5}CoCuFeNiV_{0.5}$ | A1 | A1 | CoCrFeMoV | IM+ | Sigma |
| $Al_{0.5}Cr_{0.5}Fe_2Mo_{0.15}Ni_{1.5}Ti_{0.3}$ | B2+SS | B2, A1 | $CoCrFeNb_{0.5}Ti_{0.5}$ | IM+ | Laves |
| $Al_{0.6}CrFe_4Mn_{0.5}Mo_{0.3}Ni_2Ti$ | B2+SS | B2, A1 | $CoCrFeNiSi_{0.6}$ | A1 | A1 |
| $AlCo_{0.5}CrCu_{0.2}FeMn$ | B2+SS | **A2** | $CoCr_{1.5}Fe_{1.5}NiSi_{0.2}$ | A1 | A1 |
| AlCoCrFe | B2+SS | B2, A2 | $CoCuFeMnNiV_{0.5}$ | A1 | A1 |
| $AlCoCrFeTi_{0.25}$ | B2+SS | B2, A2 | $CoFeMnNiTi_{0.5}V_{0.5}$ | IM+ | **A1** |
| $AlCoCu_{0.5}Fe$ | B2+SS | B2, A2 | CoFeMoNiTi | IM+ | Laves |
| $AlCoCuNiTi_{0.25}$ | B2+SS | B2, A1, A2 | CrCuFeMn | A1+A2 | A1, A2 |
| $AlCo_2CrCuNi_3V$ | A1 | A1 | $CrCuFeMnNiTi_{0.3}$ | A1+A2 | A1, A2 |
| $AlCrCuFeNiSi_{0.25}$ | B2+SS | B2, A2 | CrMoTiV | A2 | A2 |
| $AlCrMoNi_3W_{0.5}$ | B2+SS | **A1, A2** | CrNbNiTiZr | IM+ | Laves |
| AlCuFeNi | B2+SS | B2, A1, A2 | $Cr_2FeNiTi$ | IM+ | Laves, A2 |
| $Al_2CoNb_{0.2}Ni$ | B2+SS | **Laves, B2, A2** | $CuFeMnNiTi_2$ | IM+ | Laves, A2 |
| $Co_{0.2}TaTiV$ | A2 | A2 | CuFeMnNiV | A1+A2 | **Sigma, A1** |
| $CoCr_{0.3}Cu_{0.2}FeNiV_{0.5}$ | A1 | A1 | $Hf_{0.5}NbTaW_{0.5}Zr$ | A2 | A2 |
| $CoCr_{0.5}Fe_2NiTi_{0.25}$ | A1 | A1 | HfNbTaZr | A2 | A2 |

To summarize, a fast and accurate HEA phase prediction method is presented. It is based solely on binary phase diagrams for which there exist plentiful and easily accessible data. However, there are some limitations. Certain rare phases in HEAs are not predicted due to their small dataset size. Similar to other thermodynamic prediction methods such as CALPHAD, preparation methods producing metastable phases, such as ball milling and sputtering, cannot be predicted with this method.

In closing, several remarks can be made. In order to increase the accuracy and expand the phase prediction capabilities, additional physical features need to be identified and added into the current ML method. The current method can serve as high-throughput screening to accelerate the computation-intensive methods. For example, other methods such as CALPHAD can conduct an in-depth study on these systems for detailed information about the phase transition under different temperatures or the precise control of secondary phase precipitation by fine adjustment of the composition. Thermodynamic parameters obtained from CALPHAD will also contribute to the ML prediction process. Additionally, active learning has shown promising results in designing various materials [118, 125], and it can be applied to improve this method.

## Section 4: HEA Properties

Sections 2 and 3 have presented some promising approaches for uncovering the HEA phase-composition relationships in the complex composition landscape. As for material properties, predictive calculations will be very helpful for understanding specific HEA properties such as mechanical, electrical, thermal, magnetic, and chemical. Additionally, they are apt at quantifying the controlling parameters such as elastic moduli and hardness, Curie temperature, electrical conductivity, thermal conductivity, and electron transfer. By integrating this knowledge with the data-driven methods (Sections 2, 3, and 5), it can provide a promising framework for designing the desired compositions and properties. Within this framework, it is understood that there exist certain parameters or features that can impact the structural and functional properties. These parameters or features, also called physics-based features, should be identified and included appropriately in the prescribed models in order to be able to synthesize the desired alloys and simultaneously optimize their physical properties. Two examples of this effort will be presented in the latter part of this section.

The data-driven methods have provided us with powerful tools for designing HEAs, however, we must not forget the founding principles of HEAs and their central roles. The HEA principles embody the entropic effects. In this context, several often-mentioned HEA entropic effects [131–135] are believed to influence material properties in various ways. These effects, known as the four core effects of HEAs, are listed in Table 3 to illustrate their translations to materials research. Some specific examples of the core effects on material properties are discussed below. These core effects may be considered in the compositional design of structural and functional HEAs.

Table 3. High-entropy alloy core effects and their translations to materials research.

| HEA Core Effect | Translation to Material Research |
|---|---|
| 1. Entropy-stabilized solid solutions and extended solubility | • Large alloy design space creates composition for a desirable outcome<br>• Enables tunability of other core effects |
| 2. Sluggish or anomalous diffusion kinetics | • Multi-scale microstructure enhances the material design |
| 3. Strong lattice distortion | • Enhances mechanical, radiation, and thermal resistance |
| 4. Cocktail effect | • Synergistic outcome |

The original HEAs were based on the compositional complexity and chemical homogeneity of multi-component solid solution alloys. On the other hand, the HEA concept also advocates the use and control of heterogeneous microstructure by exploiting sluggish diffusion kinetics [136, 137]. Compositional complexity is expected to favor enhancing magnetocaloric effect (MCE) because the high degree of chemical disorder in HEAs results in large fluctuations in the magnetic exchange coupling [138, 139], which can lead to an enhanced magnetic latent heat during cooling. This particular MCE feature could open the door for HEAs as next-generation magnetic-refrigerant materials. The homogeneous nature and kinetic stability of HEAs are responsible for the superior electrocatalytic activity and resilience to harsh temperature and oxidation of HEA nanoparticles based on noble metals, e.g., PtPdRhRuCe [23]. The potential of HEA catalysts can be exploited to design oxygen reduction and hydrogen evolution reactions in proton exchange membrane fuel cells. The intriguing electrochemistry of HEAs also leads to promising aqueous corrosion resistance [20] and offers design opportunities to control passivation, but the mechanisms of these vastly different electrochemical behaviors are not understood [140]. Some key scientific questions of interest include whether there exist complex oxides that regulate and control passivation unattainable outside HEAs, and whether phase-separated oxides or disordered oxide solutions are better corrosion resistant HEAs.

The unique compositional degrees of freedom of HEA enables band-structure engineering in thermoelectric alloys to increase the figure of merit, ZT [141, 142], as well as suppression of the thermal conductivity, $\kappa$, in entropy-stabilized oxides through charge fluctuations [25]. $\kappa$ can even reach below the minimum value usually associated with amorphous solids. In addition, lattice distortion scatters phonons that can also reduce $\kappa$. Thermal and thermoelectric properties are important in thermal management and energy harvesting. Due to the sluggish diffusion kinetics, HEAs are prone to exhibit structural heterogeneities that can be controlled and exploited. The concept of chemical SRO [143] and composites [14, 144] have resulted in an unprecedented balance of strength, ductility, and density. The latter phenomenon can be attributed to the resistance to dislocation motion by the high density of heterogeneities that form a hierarchy of planar defects, clusters/precipitates, grain-size distribution, and second phases [145, 146].

Among the HEA properties considered so far, entropy probably has the most impact on thermoelectric research. Recently, it has been reported that configurational entropy may serve as an useful performance indicator of the ZT of thermoelectric materials [27, 28]. As illustrated in Fig. 15, the configurational entropy simultaneously controls the carrier mobility in the electrical conduction and the phonon mean free path in the thermal transport. The Seebeck coefficient is essentially the average entropy transported per unit charge. The relation between the charge flow and the accompanied entropy flow is given by the Wiedemann-Franz relation. The vibrational entropy is embodied in the specific heat and temperature. On the other hand, entropy creation (e.g., the Joule effect) would make a thermoelectric process thermodynamically irreversible. Hence, pursuing high ZT in a material is no more than optimizing the entropy production in the thermoelectric process therein.

The above-mentioned configurational entropy-enabled core effects can be exploited to enhance the thermoelectric properties, provided that inevitable conflicts among the thermoelectric transport properties are addressed in the alloy design [28]. Upon surveying a number of promising thermoelectric materials from the configurational entropic perspective, the efficacy of a larger phase space for compositional optimization and stabilization of higher crystal lattice symmetry for realizing a thermoelectrically favorable band structure was demonstrated. The results of entropy influenced thermoelectric properties indicated that the configurational entropy needs to be sufficiently high to elicit one or more core effects but low enough to preserve reasonable carrier mobility. The configurational entropy is thus a means but not the goal. On the other hand, the degradation in carrier mobility in multi-element tunable HEAs may be compensated for by band convergence, and by tuning the effective mass and carrier concentrations to attain high thermoelectric performance.

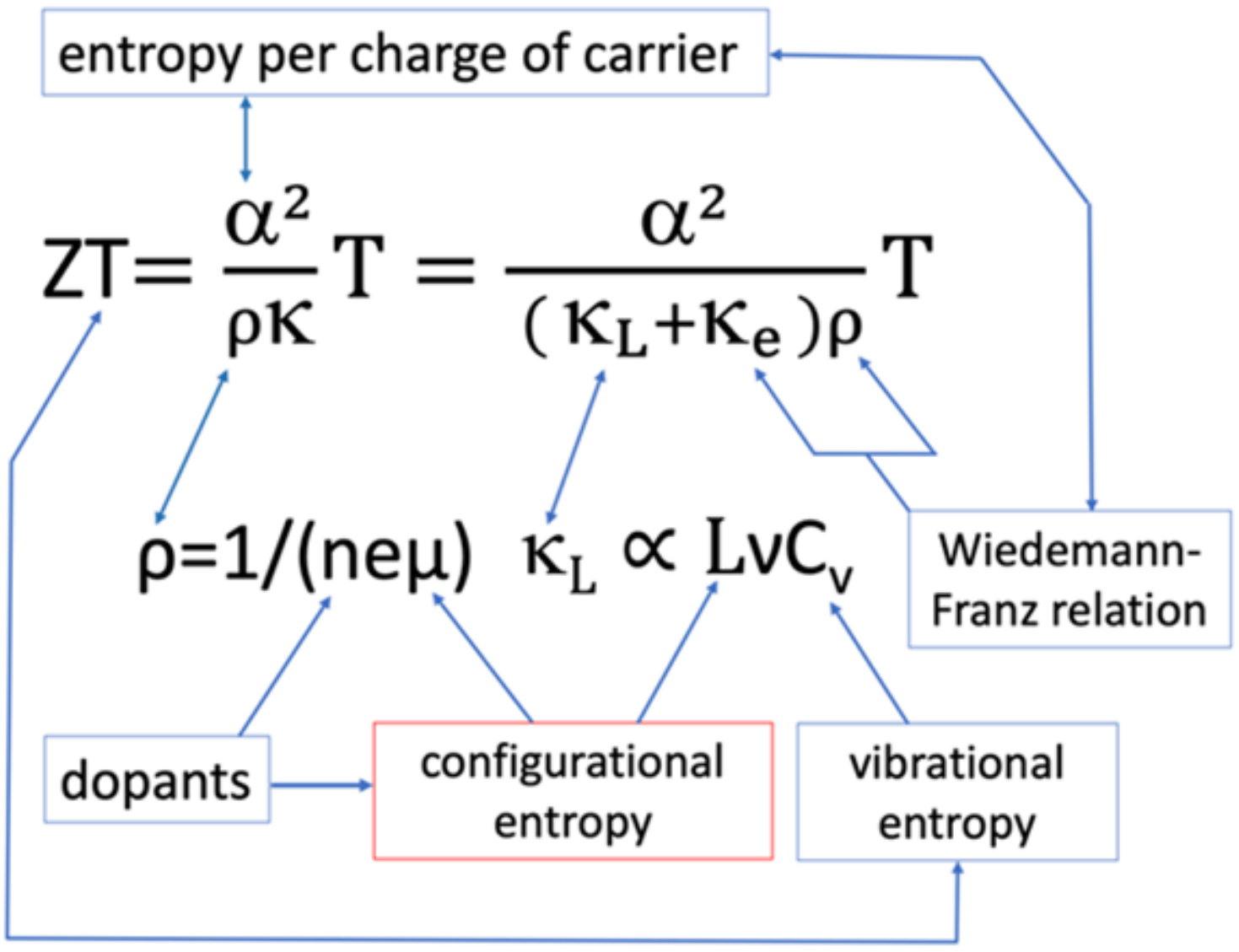

Figure 15. A schematic entropic view of thermoelectric properties. Several important mechanisms, such as the electron-phonon coupling and spin entropy, are left out for brevity. The n, μ, L, v, and $C_v$ denote the carrier concentration, carrier mobility, phonon mean free path, speed of sound, and isochoric specific heat, respectively. Figure from Poon and He [28].

As mentioned above, there are emerging efforts for using machine learning models to design HEAs with improved mechanical properties, principally hardness. Chang et al. [38] used the ANN ML algorithm to predict the hardness of HEAs and find new compositions with optimized hardness. Ninety-one HEAs containing Al, Co, Cr, Cu, Fe, Ni, Mn, and Mo with hardness data reported were contained in the database. The solid density, hardness, and atomic mass of each element, weighted by the atomic percentage of that element, were the features input in ANN. The ML model showed great agreement between the predicted and the experimental hardness results, with a value of 0.94 for the Pearson correlation coefficient. For designing new alloys with high hardness, a simulated annealing algorithm was adopted to change the composition systematically for finding global maximum hardness. HEAs designed from this model showed improvements in hardness. A general trend that the hardness increases for the same alloy system when the phase transforms from FCC to FCC+BCC to BCC+B2 is found.

Wen et al.

[39] developed a robust ML method of making HEAs with high hardness. A radial basis function kernel (svr.r) ML model was used. ML features were the HEA compositions together with the empirical parameters e/a, modulus mismatch, and the sixth square of the work function defined by the author. The dataset was composed of 155 AlCoCrCuFeNi HEA systems and their hardness values. As can be seen in Fig. 16a, the method gave reliable hardness prediction results. After that, an iterating process was used to find the HEA with the highest hardness. In each iteration, ML could sample the HEA composition space and find three new HEAs with the highest predicted hardness. The hardness values of the new HEAs were then measured experimentally and added into the training dataset for the next iteration of ML. After seven iterations, 21 new HEAs with high hardness were obtained. Their hardness values compared to the original 155 systems were plotted in Fig. 16b. The HEA $Al_{47}Co_{20}Cr_{18}Cu_5Fe_5Ni_5$ with the highest hardness was obtained in the fourth iteration. Its hardness was 883 HV, which was 14 % higher than the highest hardness value of 775 HV in the original training data.

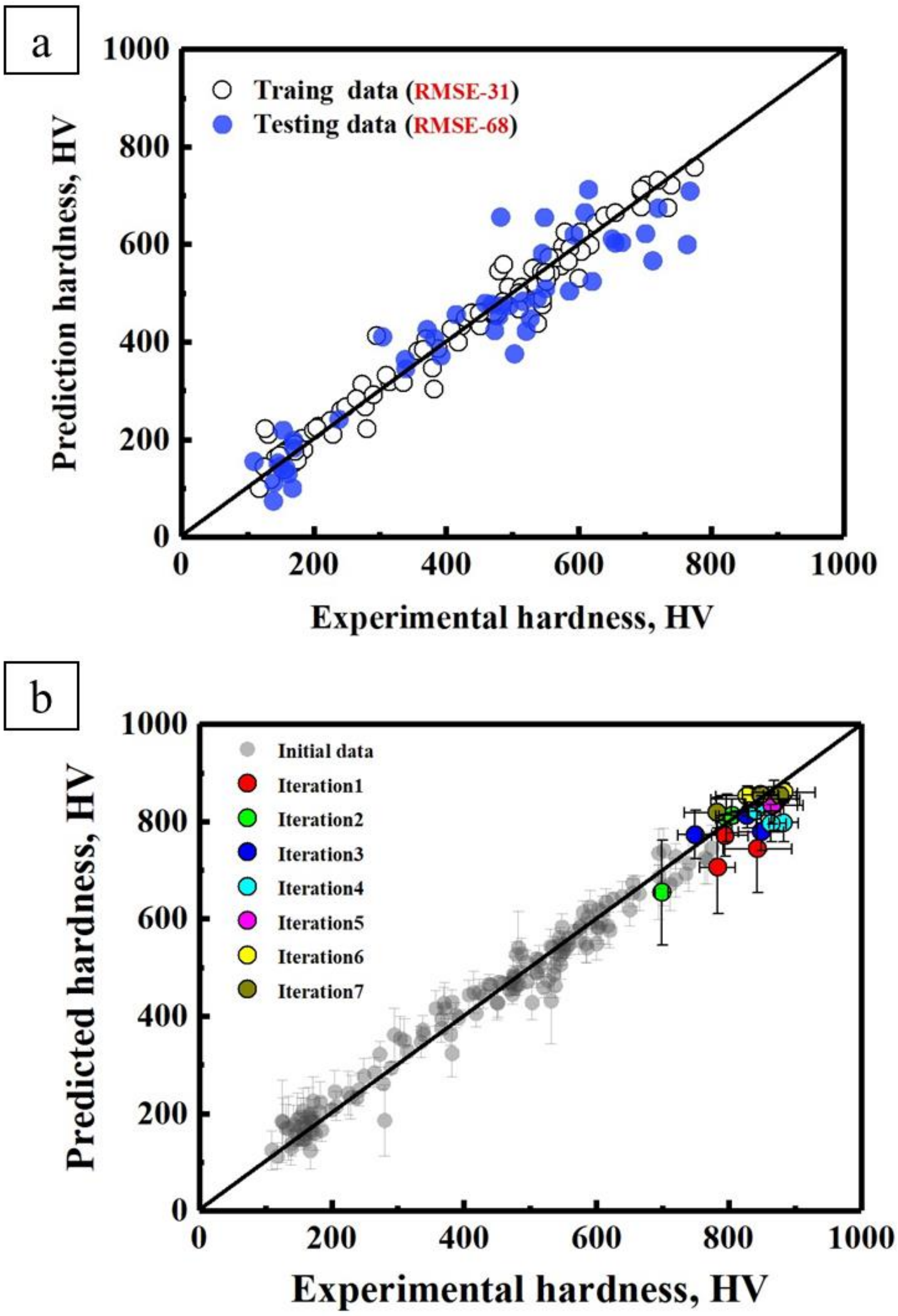

Figure 16. Plots from Wen et al [39]. (a) Comparison between the predicted and the experimental hardness values of the HEAs in both training and test dataset. (b) The predicted hardness values versus the measured values for the alloys of the original 155 training data and those synthesized in successive seven rounds of iterations.

## Section 5: Challenges, Opportunities, and Future Directions

The data-driven method is most effective when data are abundant. However, in application areas such as materials discovery, data are often limited due to large search space and resource constraints. To expand the database, fast generation of additional, high-quality data, such as from automated, high-throughput testing, is needed. Additionally new ML algorithms can be developed to enable intelligent, active learning to guide simulations and experiments to generate useful data. Additionally, physics-based constraints can also be utilized to constrain data-driven models in the low-data areas. Even with these advanced machine learning approaches, however, due to the enormous number of compositional combinations coupled with the myriad of possible phase compositions and microstructures, it must be essential to go beyond the classical machine learning methods in order to design HEAs. For example, it has been pointed out recently that designing alloys with certain phase constitution, which involves finding the thermodynamic conditions and compositions, and specifying by the desired properties, belongs to a class of constrained satisfaction problems (CSP) [147, 148]. Unfortunately, there is no general solution to this class of problems. In the future development, one can envision exploring quantum speedup, as in quantum-enhanced learning [149], to enable the training of neural networks with more complex topology. Such enhanced learning capability is needed for dealing with the joint probability distribution of all the variables of interest to HEAs design. Before such an enhancement is developed, classical ML will continue to be used as the design tool.

As described in the above sections, the data-driven approaches discussed so far have focused on predicting the formation of common HEA phases and utilizing these phase predictions to enhance HEA properties. Future research will need to address predictions of rarer phases and exotic alloy systems. The phase predictions of FCC, BCC, mixed FCC+BCC, and the most common IM phases, are accurate from existing methods. However, for less common IM phase predictions, further work is needed. Different factors influence the formation of one IM phase over another phase. Each IM phase requires a judicious study to determine the relevant controlling parameters. This has motivated the use of data and knowledge motivated adaptive features for a given type of intermetallic alloy.

Here, to demonstrate this, a simple model of the Heusler phase [150] prediction with regards to HEAs is presented. Heusler alloys, possessing the $L2_1$ structure, have general composition $X_2YZ$, where the chemical symbols X, Y, and Z are limited to certain elements [150]. The addition of Heusler alloys to a HEA has the potential to increase its mechanical properties [151]. The Heusler phase has a superior creep resistance compared with the B2 phase due to limited slip [152, 153].

Of the known HEAs with a Heusler phase present, $Ni_2TiAl$ accounts for the majority of current studied Heusler phases formed in HEAs. While there currently is a nontrivial number of known HEAs with a Heusler phase present, as shown in Table 4, a prediction model for Heusler phase formation in HEAs is lacking. A prediction model based on CALPHAD would simulate the phase content in an equilibrium solution, but this would be difficult to implement with the phase formation of a Heusler alloy since they occur typically on grain boundaries in HEAs due to a segregation-induced phase transition [154].

Table 4. HEAs containing a Heusler phase are listed. The columns contain the alloy system, the composition of the HEA, the preparation method: AC=As-cast, WQ=Water quenched, CR=Cold rolled, phases present excluding the Heusler phase, and the referenced source. For the cold rolling, the thickness reduction is included in the parenthesis. For the annealing process, the annealing temperature (C) and duration (min, h, d) are listed. All notations are in Strukturbericht designations except the phases FCC, BCC, HCP, Sigma, $\chi$, and $\eta$, which correspond to Strukturbericht designations A1, A2, A3, $D8_b$, A12, and $D0_{24}$. Laves phase corresponds to C14, C15, or C36.

| **Alloy system** / **Composition** | **Preparation Methods** | **Non-$L2_1$ Phases** | **Ref.** |
|---|---|---|---|
| **AlBCoCrFeNiTi** | | | |
| $Al_6B_{0.1}Co_{30}Cr_{15}Fe_{13}Ni_{29.9}Ti_6$ | AC + 1165 C / 2 h + CR (-65 %) + 1165 / 2 min + 800 C / 720 h | FCC + $L1_2$ | [155] |
| | AC + 1165 C / 2 h + CR (-65 %) + 1165 / 2 min + 900 C / 720 h | FCC + $L1_2$ | [155] |
| **AlCoCrCuFeNiTi** | | | |
| $AlCo_{0.5}CrCu_{0.5}FeNi_{1.5}Ti_{0.4}$ | AC | BCC | [156] |
| $Al_{0.3}CoCrCu_{0.3}FeNiTi_{0.2}$ | AC + 1150 C / 1 h + CR (-70 %) + 1150 C / 5 min | FCC + B2 | [157] |
| | AC + 1150 C / 1 h + CR (-70 %) + 1150 C / 5 min + 600 C / 150 h | FCC + B2 + Sigma + $L1_2$ + BCC | [157] |
| | AC + 1150 C / 1 h + CR (-70 %) + 1150 C / 5 min + 800 C / 0.5 h | FCC + B2 + Sigma + $L1_2$ | [157] |
| **AlCoCrCuNiTiY** | | | |
| $AlCoCrCuNiTiY_{0.5}$ | AC | BCC + FCC + C15 | [158] |
| $AlCoCrCuNiTiY_{0.8}$ | AC | BCC + C15 | [158] |
| AlCoCrCuNiTiY | AC | BCC + C15 + Unknown | [158] |
| **AlCoCrFeHfNiTi** | | | |
| $Al_{9.5}Co_{25}Cr_8Fe_{15}Hf_{0.5}Ni_{36}Ti_6$ | AC + 1220 C / 20 h | FCC + $L1_2$ | [159] |
| | AC + 1140 C / 20 h | FCC + $L1_2$ | [159] |
| | AC + 1220 C / 20 h + 900 C / 50 h | FCC + $L1_2$ | [159] |
| | AC + 1220 C / 20 h + 950 C / 100 h | FCC + $L1_2$ | [159] |
| **AlCoCrFeMoNiTi** | | | |
| $Al_{9.5}Co_{25}Cr_8Fe_{15}MoNi_{36}Ti_6$ | AC + 1220 C / 20 h + 900 C / 50 h | FCC + $L1_2$ | [159] |
| | AC + 1220 C / 20 h + 950 C / 100 h | FCC + $L1_2$ | [159] |
| **AlCoCrFeNiTi** | | | |
| $Al_{0.25}CoCrFeNiTi_{0.75}$ | AC | FCC + $\chi$ | [160] |
| $Al_{10}Co_{25}Cr_8Fe_{15}Ni_{36}Ti_6$ | AC + 1220 C / 20 h + 900 C / 50 h | FCC + $L1_2$ | [159] |
| | AC + 1220 C / 20 h + 900 C / 50 h + Bridgman process | FCC + $L1_2$ | [159] |
| | AC + 1220 C / 20 h + 950 C / 100 h | FCC + $L1_2$ | [159] |
| $Al_{12}Co_{20}Cr_{17}Fe_{35}Ni_{12}Ti_4$ | AC | FCC + BCC + B2 | [161] |
| $Al_4Co_{23.5}Cr_{23.5}Fe_{23.5}Ni_{23.5}Ti_2$ | AC + CR (-30 %) + 1000 C / 2 h + 800 C / 18 h + WQ | FCC + $L1_2$ | [162] |
| | AC + CR (-30 %) + 650 C / 4 h + WQ | FCC + $L1_2$ | [162] |
| | AC + CR (-30 %) + 1000 C / 2 h | FCC + $L1_2$ | [163] |
| | AC + 1200 C / 4 h + WQ + CR (-30 %) + 1000 C / 2 h + 800 C / 18 h | FCC + $L1_2$ | [164] |
| | AC + CR (-30 %) + 1000 C / 2 h + 700 C / 18 h | FCC + $L1_2$ | [163] |
| | AC + CR (-30 %) + 1000 C / 2 h + 750 C / 18 h | FCC + $L1_2$ | [163] |
| | AC + CR (-30 %) + 1000 C / 2 h + 800 C / 18 h | FCC + $L1_2$ | [163] |
| | AC + CR (-30 %) + 1000 C / 2 h + 850 C / 18 h | FCC + $L1_2$ | [163] |
| | AC + CR (-30 %) + 1000 C / 2 h + 900 C / 18 h | FCC | [163] |
| | AC + CR (-30 %) + 1000 C / 2 h + 800 C / 0.5 h | FCC + $L1_2$ | [163] |
| | AC + CR (-30 %) + 1000 C / 2 h + 800 C / 8 h | FCC + $L1_2$ | [163] |
| | AC + CR (-30 %) + 1000 C / 2 h + 800 C / 48 h | FCC + $L1_2$ | [163] |
| $Al_4Co_{23.75}Cr_{23.75}Fe_{23.75}Ni_{23.75}Ti$ | AC + CR (-30 %) + 1000 C / 2 h | FCC + $L1_2$ | [163] |
| $Al_{6.25}Co_{17.5}Cr_{26.25}Fe_{26.25}Ni_{17.5}Ti_{6.25}$ | AC | BCC + FCC | [151] |
| $Al_{6.25}Co_{17.5}Cr_{35}Fe_{17.5}Ni_{17.5}Ti_{6.25}$ | AC | BCC + Sigma | [151] |
| $Al_6Co_{22.75}Cr_{22.75}Fe_{22.75}Ni_{22.75}Ti_3$ | AC + CR (-30 %) + 1000 C / 2 h | FCC + $L1_2$ | [163] |
| $Al_{8.3}Co_{17.5}Cr_{26.25}Fe_{26.25}Ni_{17.5}Ti_{4.2}$ | AC | BCC + FCC | [151] |
| $Al_{8.3}Co_{17.5}Cr_{35}Fe_{17.5}Ni_{17.5}Ti_{4.2}$ | AC | BCC | [151] |
| $Al_{9.4}Co_{17.5}Cr_{26.25}Fe_{26.25}Ni_{17.5}Ti_{3.1}$ | AC | BCC + FCC | [151] |
| $Al_{9.4}Co_{17.5}Cr_{35}Fe_{17.5}Ni_{17.5}Ti_{3.1}$ | AC | BCC | [151] |
| $Al_9Co_{22}Cr_{22}Fe_{22}Ni_{22}Ti_3$ | AC + CR (-30 %) + 1000 C / 2 h | FCC + $L1_2$ | [163] |
| **AlCrFeMnNi** | | | |
| $Al_{7.5}Cr_6Fe_{40.4}Mn_{34.8}Ni_{1.3}$ | AC + 500 C / 13 d or 42 d | FCC | [165] |
| $Al_4Cr_{15}Fe_{33.5}Mn_{10}Ni_{33.5}Ti_4$ | AC + 1150 C / 2 h + CR (-66 %) + 1150 C / 2 min + 800 C / 1 h | FCC + $L1_2$ + η | [154] |
| $Al_5Cr_{15}Fe_{33.5}Mn_{10}Ni_{33.5}Ti_3$ | AC + 1150 C / 2 h + CR (-66 %) + 1150 C / 2 min + 800 C / 1 h | FCC + $L1_2$ | [154] |
| $Al_{12}Cr_{17}Fe_{35}Mn_{20}Ni_{12}Ti_4$ | AC | BCC | [161] |
| **AlCrFeMnTi** | | | |
| $AlCrFeMnTi_{0.25}$ | AC | BCC | [166] |
| $Al_{1.5}CrFeMnTi$ | AC | BCC + C14 | [166] |
| | AC + (750, 850, or 1200) C / 168 h + WQ | BCC + C14 | [75] |
| | AC + 1000 C / 504 h + WQ | BCC + C14 | [75] |
| $Al_2CrFeMnTi$ | AC | BCC + C14 | [166] |
| $Al_2CrFeMnTi_{0.25}$ | AC | BCC | [166] |
| $Al_3CrFeMnTi_{0.25}$ | AC | BCC + A12 + A7 | [166] |
| $Al_4CrFeMnTi_{0.25}$ | AC | BCC + A12 + A7 | [166] |
| **AlCrFeNiTi** | | | |
| $Al_5Cr_{32}Fe_{35}Ni_{22}Ti_6$ | AC | BCC + FCC | [167] |
| | AC + 1100 C / 6 h + WQ | BCC + FCC | [167] |
| | AC + 1100 C / 6 h + WQ + 700 C / 100 h | BCC + FCC + Sigma | [167] |
| | AC + 1100 C / 6 h + WQ + 800 C / 100 h | BCC + FCC + Sigma + η | [167] |
| | AC + 1100 C / 6 h + WQ + 900 C / 100 h | BCC + FCC + Sigma + η | [167] |
| **AlCrFeNiTiV** | | | |
| $Al_{0.5}CrFeNiTiV$ | AC + 700 C / 20 h | C15 | [64] |
| **AlFeMnNi** | | | |
| $Al_{25}Fe_{35}Mn_{25}Ni_{15}$ | AC | B2 | [168] |
| | AC + 550 C / 22 h | B2 | [168] |
| $Al_{30}Fe_{30}Mn_{20}Ni_{20}$ | AC | B2 | [169] |
| | AC + 550 C / (0.5, 12, or 72) h | B2 | [169] |
| **CoFeMnNiSn** | | | |
| CoFeMnNiSn | AC | FCC + BCC | [170] |

Some HEAs contain elements that can potentially form more than one Heusler phase. In some cases, a different intermetallic (IM) phase may form with or without the Heusler phase. Based on our knowledge, we will consider the conditions that may favor the formation of a particular Heusler phase. Based on these Heusler phase formation conditions, three prospective adaptive features are proposed:

1. The electronegativity ($\chi$) ratio defined herein as $\frac{c_{\chi_{Max}} \times \chi_{Max}}{c_{\chi_{Min}} \times \chi_{Min}}$, where Max and Min represent maximum and minimum values, respectively, indicates the strong (weak) tendency to form intermetallic phase if the electronegativity ratio is high (low).

2. More than one IM phase could form. The mixing enthalpy of Heusler phase ($\Delta H_{mix-L2_1}$) must be compared with the most negative binary in the HEA ($\Delta H_{mix-binary}$), which suggests that the ratio $\Delta H_{mix-L2_1} / \Delta H_{mix-binary}$ can serve as a feature.

3. The total atomic percentage of X, Y, and Z in the HEA is assumed to infer the tendency of forming the Heusler phase. A low total concentration could favor dissolution instead of precipitation. Thus, the total X, Y, and Z concentration ($\sum c_{L2_1}$) is a good feature candidate.

4. If the concentration of one of the three Heusler phase forming elements is low relative to $\sum c_{L2_1}$, the entropy is decreased and there will be a stronger tendency to precipitate the Heusler phase instead of forming a single-phase HEA. This relative concentration ratio $c_{L2_1-min} / \sum c_{L2_1}$ can serve as an adaptive feature.

For ML training, the currently known 41 HEAs that contain the Heusler phase ($HEA_{L2_1}$) were used as the primary dataset. A comparison dataset of 98 HEAs that do not contain the Heusler phase ($HEA_{non-L2_1}$) was also used. HEAs in the comparison dataset are stipulated to include elements that can form a Heusler phase. The two datasets contain only annealed samples, in order to avoid the suppression of Heusler formations from rapid cooling during sample preparation. In Fig. 17, a plot with three of the four defined adaptive features as the axes is constructed. All the points are plotted based on their feature values. The two types of HEAs show significate separation. The results, Table 5, of ML using the Random Forest as classification algorithm returns prediction success rates near 75 % and 84 % for $HEA_{L2_1}$ and $HEA_{non\text{-}L2_1}$, respectively.

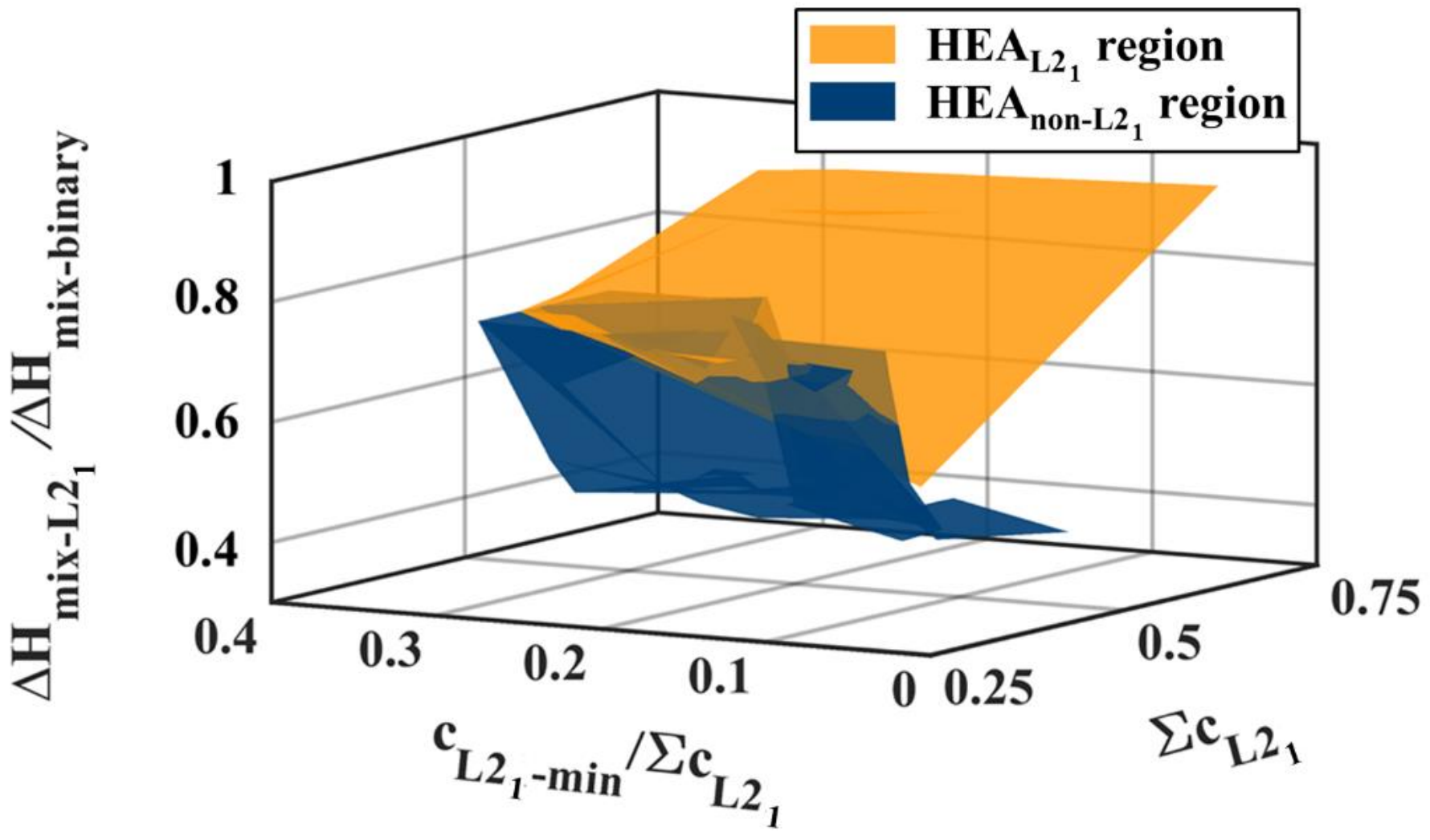

Figure 17. Three ML features plot showing distributions of $HEA_{L2_1}$ and $HEA_{non\text{-}L2_1}$.

Table. 5 The ML success rates of $HEA_{L2_1}$ and $HEA_{non\text{-}L2_1}$ predictions based on the percent of total database used as a training percent set.

| **Training %** | **$HEA_{L2_1}$ Success Rate (%)** | **$HEA_{non\text{-}L2_1}$ Success Rate (%)** |
|---|---|---|
| 90 | 77 | 84 |
| 80 | 76 | 84 |
| 75 | 76 | 82 |
| 66 | 72 | 82 |
| 50 | 71 | 81 |

The outcome of this simple model suggests that tailored features can predict other IM phases as well. For example, the ordered FCC $L1_2$ phase is widely used in HEAs to achieve a balance between strength and ductility [144, 157, 162, 171–174]. However, its formation has not been studied meticulously.

Furthermore, several more advanced techniques can be applied to improve prediction accuracy. As introduced in Section 2.3, features engineering [118, 119] can expand the feature pool by mathematically manipulating the current features. The best combinations of features are then selected for ML. Prior work [118, 119] has shown that the prediction accuracies were improved by feature engineering. Active learning [39, 125, 175, 176] is another technique, especially for the problem with small databases, to improve prediction accuracy. The current majority of the HEA database has been developed through systemic studies of HEA formation in favorable elemental systems. Over 1,000 unique HEA compositions have been reported [45]. Nevertheless, the potential number of HEAs is much larger. Active learning will be crucial to explore these untapped regions. The active learning process is an iterative one consisting of gathering for a database, ML applied to the database to predict phases and properties, experimental narrowing of uncertainties of the predictions, and then refining the database. Further iterations improve the ML prediction ability and further expand the database.

Finally, as a future plan, the founding principles of high-entropy alloys mentioned in Section 4 should be brought to the fore in the formulation of ML features. Entropy is inherent in the thermodynamic parameters used in the data-driven models described, the entropy effects of HEAs nevertheless have not been beneficially utilized in these models. New models must be developed to exploit the entropy effects that are essential for understanding the fundamental factors that control the phases and their properties.

**Section 6: Summary**

To summarize, one of the primary challenges of high-entropy alloys is how to predict their phase formation and properties given the hugely complex compositional space. We have reviewed the development of various computational models that began with the use of empirical parameters to analyze high-entropy alloys formation. As the modeling efforts evolve, multiple parameters representing atomistic, thermodynamic, and chemical as well as mechanical effects were conjointly used in various machine learning models. This was found to be more efficient, and it has been utilized to further phase predictions. Meanwhile, first-principles calculations have also been found to be successful in phase and properties predictions. As featured in this article, the use of alloy phase diagrams, complemented with formation enthalpy and stoichiometry of intermetallic alloys, has provided an efficient framework for predicting the various high-entropy alloy phase domains in the complex, high-dimensional composition space. Overall, this approach, which has been validated by experiment, has achieved a prediction accuracy of higher than 80 percent for common solid solution phases and a few intermetallic phases.

Despite the success of current models, however, the founding principles of high-entropy alloys have not been widely utilized in the formulation of these models. For this reason, there is a need to discuss entropic effects on functional properties. Thermoelectric properties, which are obviously impacted by configuration entropy, are used as a case study. It was noted that the synergy of two major classes of materials: high entropy alloys and thermoelectric materials lead to the conception of "high-entropy thermoelectrics". The entropy enabled core effects can be strategically exploited to produce a net positive effect on the thermoelectric properties. Future studies should include entropic effects on mechanical and thermal properties such as yield strength, toughness, fatigue, and thermal expansion.

In order to achieve synthesizability and simultaneously optimize material properties, physics-based features must be identified and included appropriately in the prescribed models. However, as is common in materials research, the use of machine learning is often limited by the size and depth of the database and selection of relevant features for supervised learning. There exists the opportunity to develop active learning as well as data and knowledge inspired adaptive features, which not only will enhance the accuracy of alloy design, but will also expand the database to advance the machine learning models to the next level.

**Acknowledgment:**

This work is supported by the Office of Naval Research grant N00014-19-1-2420.